\RequirePackage{lineno}
\documentclass[prd,twocolumn,showpacs,preprintnumbers,floatfix,amsmath,amssymb,amsfonts]{revtex4}

\usepackage{graphicx}
\usepackage{amsmath}
\usepackage{dcolumn}
\usepackage{epsfig}
\usepackage{psfrag}
\usepackage{subfigure}
\usepackage{hyperref}
\usepackage{multirow}

\RequirePackage{xspace}

\def\stat{\mbox{$\;$(stat.)}}
\def\syst{\mbox{$\;$(syst.)}}

\def\TeV{\ifmmode {\mathrm{\ Te\kern -0.1em V}}\else
	                   \textrm{Te\kern -0.1em V}\fi}%
\def\GeV{\ifmmode {\mathrm{\ Ge\kern -0.1em V}}\else
	                   \textrm{Ge\kern -0.1em V}\fi}%
\def\MeV{\ifmmode {\mathrm{\ Me\kern -0.1em V}}\else
	                   \textrm{Me\kern -0.1em V}\fi}%
\def\keV{\ifmmode {\mathrm{\ ke\kern -0.1em V}}\else
	                   \textrm{ke\kern -0.1em V}\fi}%
\def\eV{\ifmmode  {\mathrm{\ e\kern -0.1em V}}\else
	                   \textrm{e\kern -0.1em V}\fi}%
\let\tev=\TeV
\let\gev=\GeV
\let\mev=\MeV

\def\TeVc{\ifmmode {\mathrm{\ Te\kern -0.1em V}/c}\else
	                   {\textrm{Te\kern -0.1em V}/$c$}\fi}%
\def\GeVc{\ifmmode {\mathrm{\ Ge\kern -0.1em V}/c}\else
	                   {\textrm{Ge\kern -0.1em V}/$c$}\fi}%
\def\MeVc{\ifmmode {\mathrm{\ Me\kern -0.1em V}/c}\else
	                   {\textrm{Me\kern -0.1em V}/$c$}\fi}%
\def\keVc{\ifmmode {\mathrm{\ ke\kern -0.1em V}/c}\else
	                   {\textrm{ke\kern -0.1em V}/$c$}\fi}%
\def\eVc{\ifmmode  {\mathrm{\ e\kern -0.1em V}/c}\else
	                   {\textrm{e\kern -0.1em V}/$c$}\fi}%
	
\def\cm{\ifmmode  {\mathrm{\ cm}}\else	                   
\textrm{~cm}\fi}%

\def\mm{\ifmmode  {\mathrm{\ mm}}\else	                   
\textrm{~mm}\fi}%

\def\ipb{\mbox{pb$^{-1}$}}
\def\inb{\mbox{nb$^{-1}$}}

\begin{document}

\preprint{CERN-PH-EP-2011-180}

\preprint{Submitted to Physical Review D}

\title{Measurement of $\textbf{\emph{D}}^{*\pm}$ meson production in jets from $\textbf{\emph{pp}}$ collisions \\ at $\mathbf{\sqrt{\emph{s}}}$~=~7~TeV with the ATLAS detector}

\author{The ATLAS Collaboration} 

\begin{abstract}
This paper reports a measurement of $D^{*\pm}$ meson production in jets 
from proton-proton collisions at a center-of-mass energy of $\sqrt{s}=7\tev$
at the CERN Large Hadron Collider. The measurement is based on a data sample
recorded with the ATLAS detector with an integrated luminosity of $0.30~\ipb$ for jets
with transverse momentum between $25$ and $70\gev$ in the pseudorapidity range
$|\eta|<2.5$. $D^{*\pm}$ mesons found in jets are fully reconstructed in the decay chain:
$D^{*+}\to D^0\pi^+, D^0\to K^-\pi^+$, and its charge conjugate. The production
rate is found to be $N(D^{*\pm})/N({\rm jet})=0.025\pm 0.001\stat \pm 0.004\syst$
for $D^{*\pm}$ mesons that carry a fraction $z$ of the jet momentum in the range
$0.3<z<1$. Monte Carlo predictions fail to describe the data at small values of $z$,
and this is most marked at low jet transverse momentum.
\end{abstract}

\maketitle

\section{Introduction}
\label{sec:intro}

Heavy flavor production in high-energy interactions has produced
interesting tests of quantum chromodynamics (QCD) and valuable
information on fragmentation and decay properties. Early measurements
of $b$-hadron production cross sections in high-energy $p\bar{p}$
collisions~\cite{PhysRevLett.71.500,PhysRevLett.71.2396,Abbott:1999se,Albajar1988405}
were higher than the available theoretical calculations.
A satisfactory agreement between measurement and theory was reached for
$b$-hadron production with improved analysis methods, see for
example~\cite{PhysRevLett.85.5068,PhysRevD.71.032001}, and more accurate
QCD predictions, as discussed in~\cite{Cacciari,Mangano,Cacciarietal}.
Measurements of $c$-hadron production~\cite{Albajar:1990wn,Acosta:2003ax,Abe:1989tc}
are however less conclusive and additional experimental data are needed
to probe the theory, in which nonperturbative effects such as fragmentation
have a significant impact on the theoretical calculations.
With collisions at higher center-of-mass energy at the CERN Large Hadron Collider (LHC),
the kinematical range accessible to experiment has been significantly
extended~\cite{LHCb,CMS,ATLAS,ALICE}.
New measurements of heavy flavor production will help in testing
improved QCD-based models.
Moreover, precise knowledge of heavy quark production is important for
an understanding of the backgrounds in searches for new phenomena beyond
the standard model if they include decays to heavy quarks. 

One method to study the production of heavy quarks is to measure
$D^{*\pm}$ mesons produced inside jets~\cite{Arnison:1984rj,
Albajar:1990wn,Abe:1989tc,:2009df} by fully reconstructing the decay chain:
$D^{*+}\to D^0\pi^+, D^0\to K^-\pi^+$ and its charge conjugate. This
paper reports a measurement of $D^{*\pm}$ meson production in
jets from proton-proton collisions at a center-of-mass energy of
$\sqrt{s}=7\tev$ at the LHC. The measured quantity reported here is $\cal R$,
the ratio of $D^{*\pm}$ produced in jets, hereafter denoted ``$D^{*\pm}$ jets'',  
to any type of jet, called ``inclusive jets'', as a function of the jet
transverse momentum $p_{\text{T}}$, and the ratio, $z$, of the $D^{*\pm}$ momentum
along the jet axis to the jet energy, $z = p_{||}(D^{*\pm})/E(\rm{jet})$.
${\cal R}$ is defined by
\begin{equation}
{\cal{R}}(p_{\text T},z) = \frac{N_{D^{*\pm}}(p_{\text T},z)}{N_{{\rm jet}}(p_{\text T})},
\end{equation}
where $p_{||}(D^{*\pm})$ is the momentum of the $D^{*\pm}$ meson along the
jet axis; $E({\rm jet})$ is the energy of the $D^{*\pm}$ jet; $N_{D^{*\pm}}(p_{\text T},z)$ is the
number of jets that contain a $D^{*\pm}$ meson, in the corresponding $p_{\text T}$ and
$z$ bin, and $N_{{\rm jet}}(p_{\text T})$ is the number of inclusive jets in the
$p_{\text T}$ bin. 

\section{The ATLAS Detector}

The ATLAS detector is described in detail elsewhere~\cite{Aad:2008zzm};
only the components relevant to this analysis are described here. The ATLAS
coordinate system has the origin at the nominal beam-beam interaction point. 
The azimuthal angle $\phi$ is measured around the beam axis, $z$, in the $x-y$
transverse plane, and the polar angle $\theta$ is the angle from the beam axis. 
For particles and jets, the transverse momentum is defined as $p_{\rm T} = p \sin\theta$,
where $p$ is the momentum, and the pseudorapidity is defined as $\eta=-\ln\tan\theta/2$.

The inner tracking detector (ID) has full coverage in $\phi$ and is contained
inside a central solenoid providing a 2~T magnetic field. The ID
consists of a silicon pixel detector, a silicon microstrip detector
(SCT) and a transition radiation tracker (TRT). The pixel detector and SCT cover
the pseudorapidity range $|\eta|<2.5$ and the TRT covers $|\eta| < 2.0$.
Reconstructed charged tracks traversing the central part of the detector typically have
11 silicon hits (3 pixel clusters and 8 strip clusters), and more than 30
TRT hits.
	
The calorimeter system used to reconstruct jets is placed immediately outside
the solenoid. A high granularity lead liquid-argon  electromagnetic sampling
calorimeter, with excellent energy and position resolution, covers the
pseudorapidity range $|\eta|<3.2$ (a barrel covers $|\eta|<1.475$ and 
two end-caps cover $1.375<|\eta|<3.2$). 
Two different detector technologies are used for the hadronic calorimetry.
The barrel ($|\eta|<1.0$) and extended barrel ($0.8<|\eta|<1.7$)
calorimeters are made of steel and scintillator tiles while in the
end-caps ($1.5<|\eta|<3.2$) copper and liquid-argon are used. 
Forward copper and tungsten liquid-argon calorimeters provide both
electromagnetic and hadronic measurements with coverage of $|\eta|<4.9$.
	
The ATLAS detector has a three-level trigger system: Level 1 (L1),
Level 2 (L2) and Event Filter (EF). L1 is a hardware trigger system,
while L2 and EF are software based. For the measurement described here,
the data are collected using the L1 calorimeter-based jet
trigger and a system of minimum-bias trigger scintillators
(MBTS)~\cite{Trigger}. The L1 calorimeter trigger uses coarse detector
information to identify areas in the calorimeter with energy deposits
above a certain threshold. A simplified jet finding algorithm based
on a sliding window of configurable size is used to trigger events.
This algorithm uses towers with a granularity of
$\Delta\phi\times\Delta\eta=0.2\times0.2$ as inputs. The MBTS consist of
32 scintillator counters that are each $2\cm$ thick, organized into two
disks perpendicular to the beam located at $z=\pm3.56$~m. This leads to
a coverage of $2.09<|\eta|<3.84$. The MBTS trigger is configured to require
at least one hit above threshold from each side of the detector.

\section{Data and Monte Carlo Samples}
\label{sec:sample}

The analysis uses an integrated luminosity of
$0.30~\ipb$, measured with an error of $3.4\%$~\cite{Aad:2011dr,lumi},
recorded between April and July 2010 with
three L1 calorimeter jet triggers, requiring a transverse energy
associated to the L1 jet above 5, 10 and 15~GeV, respectively.
Because of the increase in instantaneous luminosity during the data
taking period, the 5 and 10~GeV triggers were progressively prescaled.
In the subsequent data taking periods the prescale factors of low-threshold
jet triggers became too prohibitive to extend the analysis to higher
integrated luminosity.

To validate the Monte Carlo (MC) simulation of the jet trigger efficiency,
a data sample collected with the MBTS trigger is used, whose integrated
luminosity corresponds to approximately $1~\inb$ after taking into account
its prescale factor.

The MC simulated events used for the correction of the signal yield for
detector effects are produced with the 
{\footnotesize{PYTHIA}} 6.421 event generator~\cite{Sjostrand:2006za}.
It implements leading-order (LO) perturbative QCD (pQCD) matrix elements
for $2\to 2$ processes, $p_{\rm T}$-ordered parton showers calculated in a
leading-logarithmic approximation, an underlying event simulation using
multiple-parton interactions, and uses the Lund string model for 
hadronization. The Martin-Stirling-Thorne-Watt
LO proton structure functions~\cite{Sherstnev:2007nd,
Martin:2009iq} with the generator tune described
in Ref.~\cite{ATLAS-PHYS-PUB-2010-002} are used for the generation of the MC
sample. The generated samples are passed through a full
simulation~\cite{:2010wqa} of the ATLAS detector and trigger based on
{\footnotesize{GEANT4}}~\cite{Agostinelli:2002hh}. Finally, the simulated
events are reconstructed and selected using identical procedures as for
the data.

The measured $D^{*\pm}$ jet production rates are compared to
predictions from different MC generators: {\footnotesize{PYTHIA}},
described above, and {\footnotesize{HERWIG}}~6~\cite{Corcella:2000bw},
with the AUET1 generator tune described in Ref.~\cite{ATL-PHYS-PUB-2010-014},
which also employs LO pQCD matrix elements, but uses an angle-ordered
parton shower model and a cluster hadronization model. The underlying
event for the {\footnotesize{HERWIG}}~6 samples is generated using
the {\footnotesize{JIMMY}}~\cite{Jimmy} package to model multiple-parton interactions.
Further comparison of the measurement is performed to the
next-to-leading-order (NLO) pQCD calculation implemented in
{\footnotesize{POWHEG}}~\cite{Nason:2004rx, Frixione:2007vw, Alioli:2010xd,
Alioli:2010xa}. The CTEQ~6.6~\cite{Nadolsky:2008zw} parametrization
is chosen as the parton density function of the proton.
In order to compare with data at the particle level,  
nonperturbative corrections have to be applied. This is done using
leading-logarithmic parton shower MC programs: the {\footnotesize{PYTHIA}}
and {\footnotesize{HERWIG}} models introduced above.
$D^{*\pm}$ mesons are produced either directly in the fragmentation
of charm quarks or via a cascade in the fragmentation of bottom quarks
and the subsequent decay of a $b$ hadron.
The charm and bottom quark masses are set to $1.5\,\gev$ and $4.75\,\gev$,
respectively. The fraction of charm quarks that fragment to a $D^{*\pm}$
meson, $f(c\to D^{*+})$, is set to
$0.224 \pm 0.028$~\cite{CLEO_1988,ARGUS,BABAR,CLEO_2004,BELLE};
the fraction of $b$ hadrons that decay to a final state with a $D^{*\pm}$
meson, $f(b\to D^{*+}X)$, is $0.17 \pm 0.02$~\cite{OPAL}.

\section{Event Selection}

Events are required to have at least one $pp$ primary vertex reconstructed
from at least five charged tracks with $p_{\rm T}>150\mev$ each. Only
vertices lying within $\pm 10\cm$ along the beam axis of the nominal
interaction point are considered. In events with multiple vertices,
the vertex with the largest $\sum p^2_{\rm T}$ of associated charged 
tracks is taken as the primary event vertex. For the values of instantaneous 
luminosity used in this analysis the average number of additional interactions
per beam crossing was small, about 0.3.

The anti-$k_{t}$ algorithm~\cite{Cacciari:2008gp} with radius parameter
$0.6$ is used to reconstruct jets from topological energy
clusters~\cite{:2010wv} assuming that
the reconstructed primary event vertex is at the origin of the jet. 
The clusters in the calorimeter are seeded by calorimeter cells with energy
$|E_{\rm cell}|>4\sigma$, where $\sigma$ is the RMS of the cell noise distribution. All
directly neighboring cells are added, then neighbors of neighbors are
iteratively added for all cells with signals above a secondary threshold
$|E_{\rm cell}|>2\sigma$.
Finally the energy in all further adjacent neighbors is added.
Clusters are split or merged based on the position
of local minima and maxima. The energies of cells of a cluster are summed to give the
cluster energy, and the clusters are treated as massless with energy
$E=\sum E_{\rm cell}$. The baseline calibration for these clusters corrects
their energy to the electromagnetic (EM)
scale~\cite{ATL-LARG-PUB-2006-003,springerlink:10.1140/epjc/s10052-010-1354-y,springerlink:10.1140/epjc/s10052-010-1508-y},
which is derived in test-beams, and properly
calibrates the energy of particles interacting electromagnetically in
the electromagnetic and hadronic calorimeters. Finally, a $p_{\rm T}$ and
$\eta$ dependent jet energy scale (JES)~\cite{ATLAS-CONF-2010-056,ATLAS-CONF-2010-055}
is applied to the jets to correct effects of hadronic shower response and detector-material
distributions. The JES is determined based on the detector simulation and
validated with extensive test beam and collision data studies. The jets used
in the measurement are required to have $25<p_{\rm T}<70 \gev$ and $|\eta|<2.5$.
Jets that are likely to have arisen from detector noise or cosmic rays are rejected~\cite{jet_cleaning}.

Candidates for $D^{*\pm}$ mesons inside jets are reconstructed in
the decay chain: $D^{*+}\to D^0\pi^+$, $D^0\to K^-\pi^+$ and its charge
conjugate. Two oppositely-charged tracks with $p_{\rm T}>1\gev$ are combined
to form a $D^0\to K^-\pi^+$ candidate and a second candidate
$\bar{D}^0\to K^+\pi^-$ with the $K$ and $\pi$ mass hypotheses swapped.
The $D^0$ ($\bar{D}^0$) candidate whose mass is within $50\mev$ of the PDG
value~\cite{Nakamura:2010zzi}, corresponding to slightly more than twice 
the measured mass resolution, is then combined with a
third track with $p_{\rm T}>0.5\,\gev$ having the same charge as the pion
to form a $D^{*+}\to D^0\pi^+$ ($D^{*-}\to \bar{D}^0\pi^-$) candidate.
To reduce the combinatorial background of uncorrelated pairs, the
$D^{*\pm}$ mesons are required to have a transverse momentum 
larger than $7.5\gev$, and the measured $D^0$ ($\bar{D}^0$) transverse decay
length is required to be greater than zero. The transverse decay length
is defined as $L_{xy}=\vec{r}\cdot \vec{p}_{\rm T}/|\vec{p}_{\rm T}|$, where
$\vec{r}$ is the displacement vector pointing to the $D^0$ ($\bar{D}^0$)
decay vertex from the primary vertex in the transverse plane, and the
$\vec{p}_{\rm T}$ is the transverse momentum of the $D^0$ ($\bar{D}^0$)
candidate. The $D^0$ ($\bar{D}^0$) decay vertex is obtained extrapolating the $K$ and $\pi$ tracks.
The MC simulation predicts that the selection $L_{xy}>0$ rejects
half of the combinatorial background and retains $89\%$ of the
signal, consistent with what has been observed in the data. 

The reconstructed $D^{*\pm}$ candidates are matched with the reconstructed jets
in the event. A jet is considered as a $D^{*\pm}$ jet candidate if the $D^{*\pm}$
direction is in a cone of $\Delta R=\sqrt{(\Delta\eta)^2+(\Delta\phi)^2}=0.6$
centered on the jet axis, the same value as the radius parameter of the
anti-$k_{\rm t}$ jet algorithm.
The momentum fraction $z$ of the $D^{*\pm}$ jet candidates is required
to be larger than $0.3$ due to the low reconstruction efficiency and large
combinatorial background for $D^{*\pm}$ jets with $z<0.3$.

\begin{figure}[!htb]
\begin{center}
\includegraphics[width=0.235\textwidth]{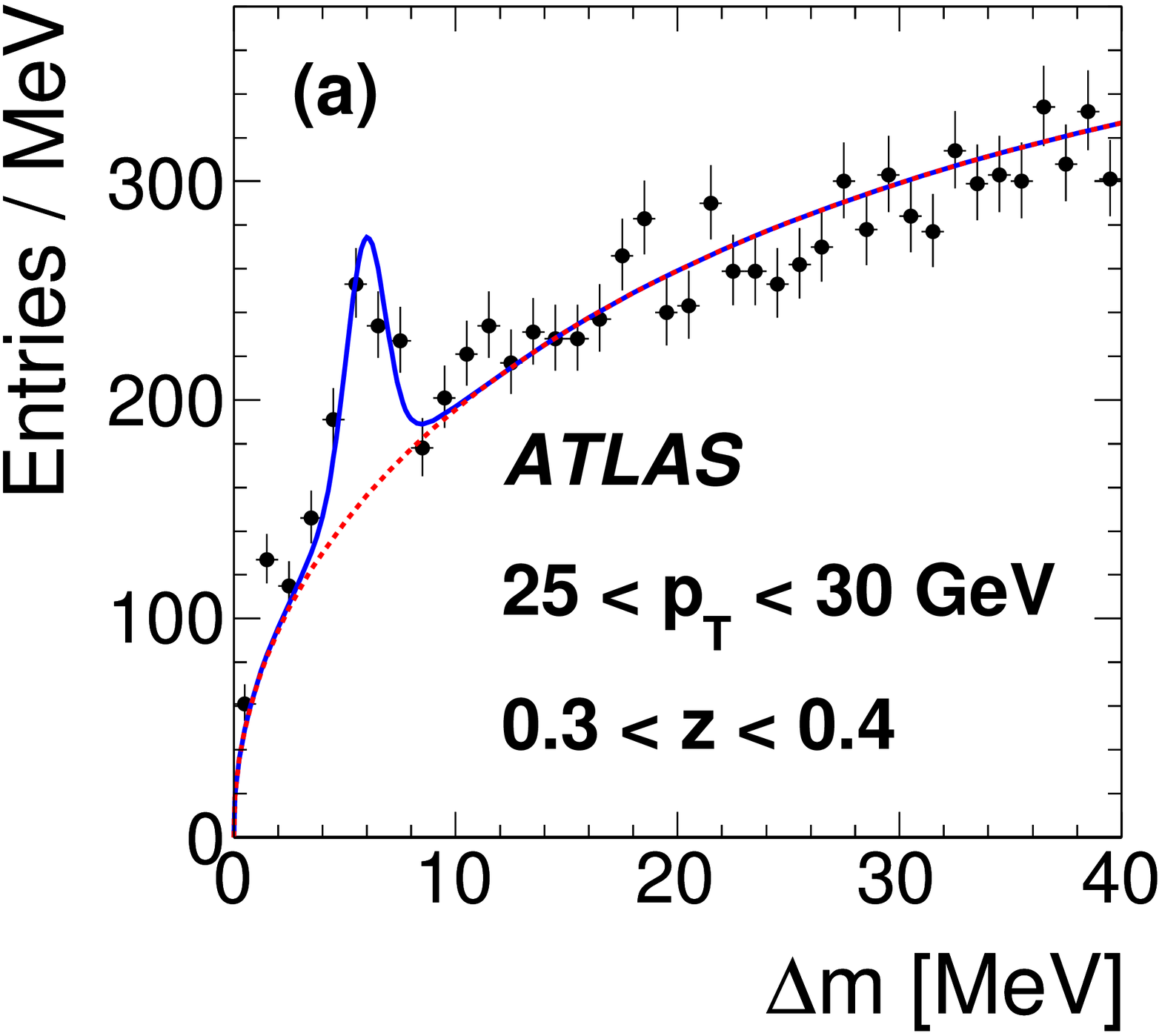}\hspace{1mm}
\includegraphics[width=0.235\textwidth]{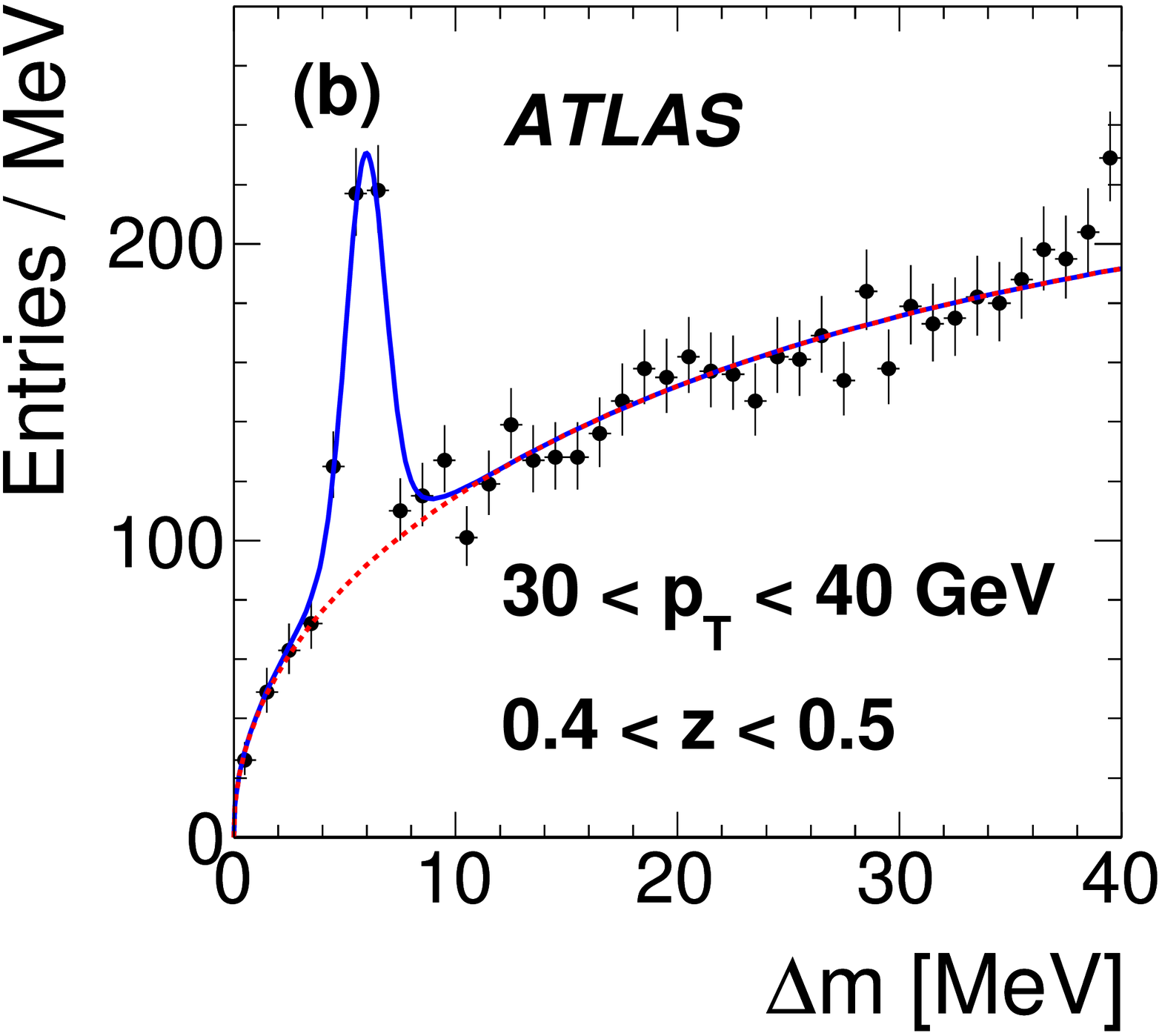}\\\vspace{1mm}
\includegraphics[width=0.235\textwidth]{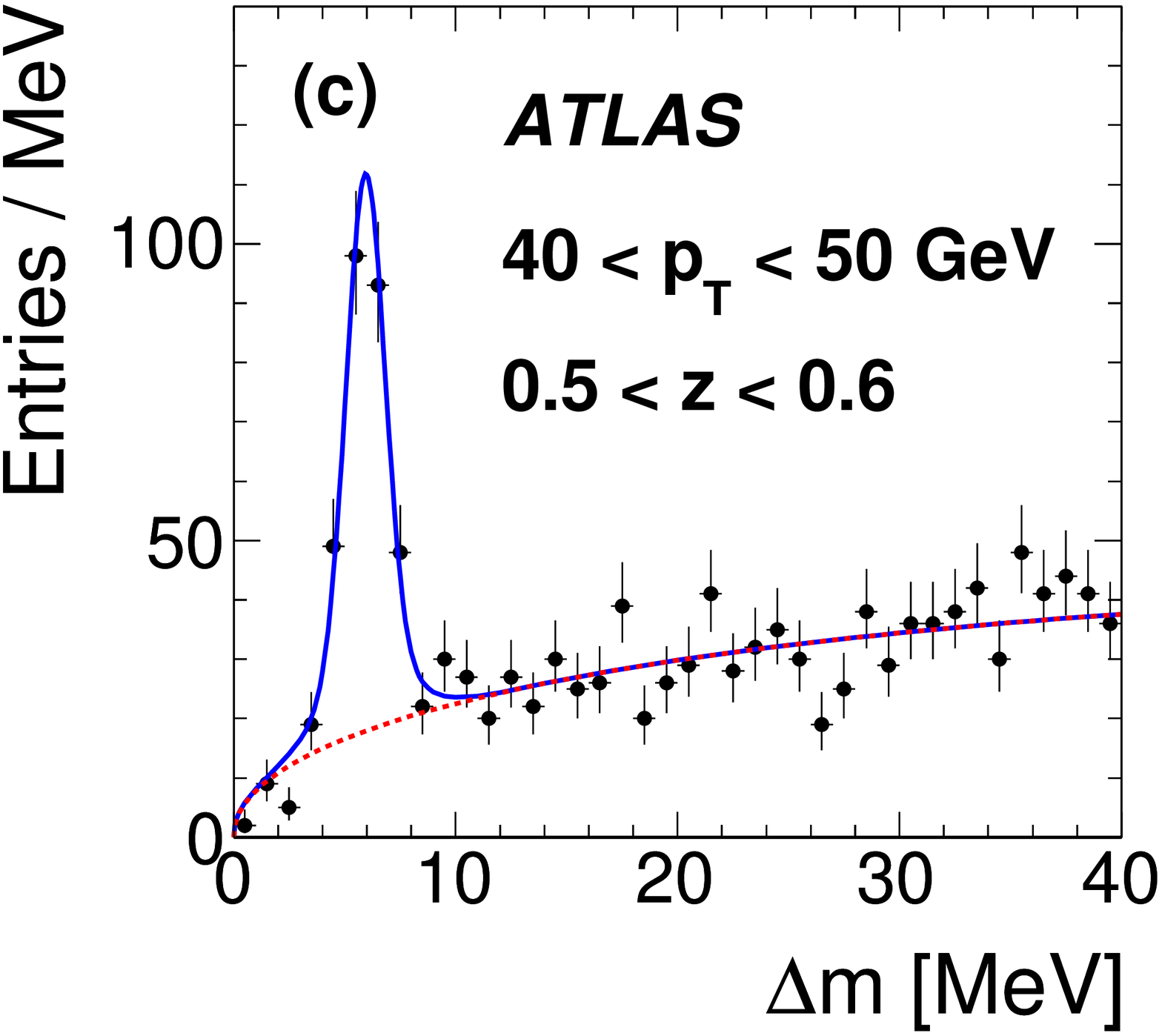}\hspace{1mm}
\includegraphics[width=0.235\textwidth]{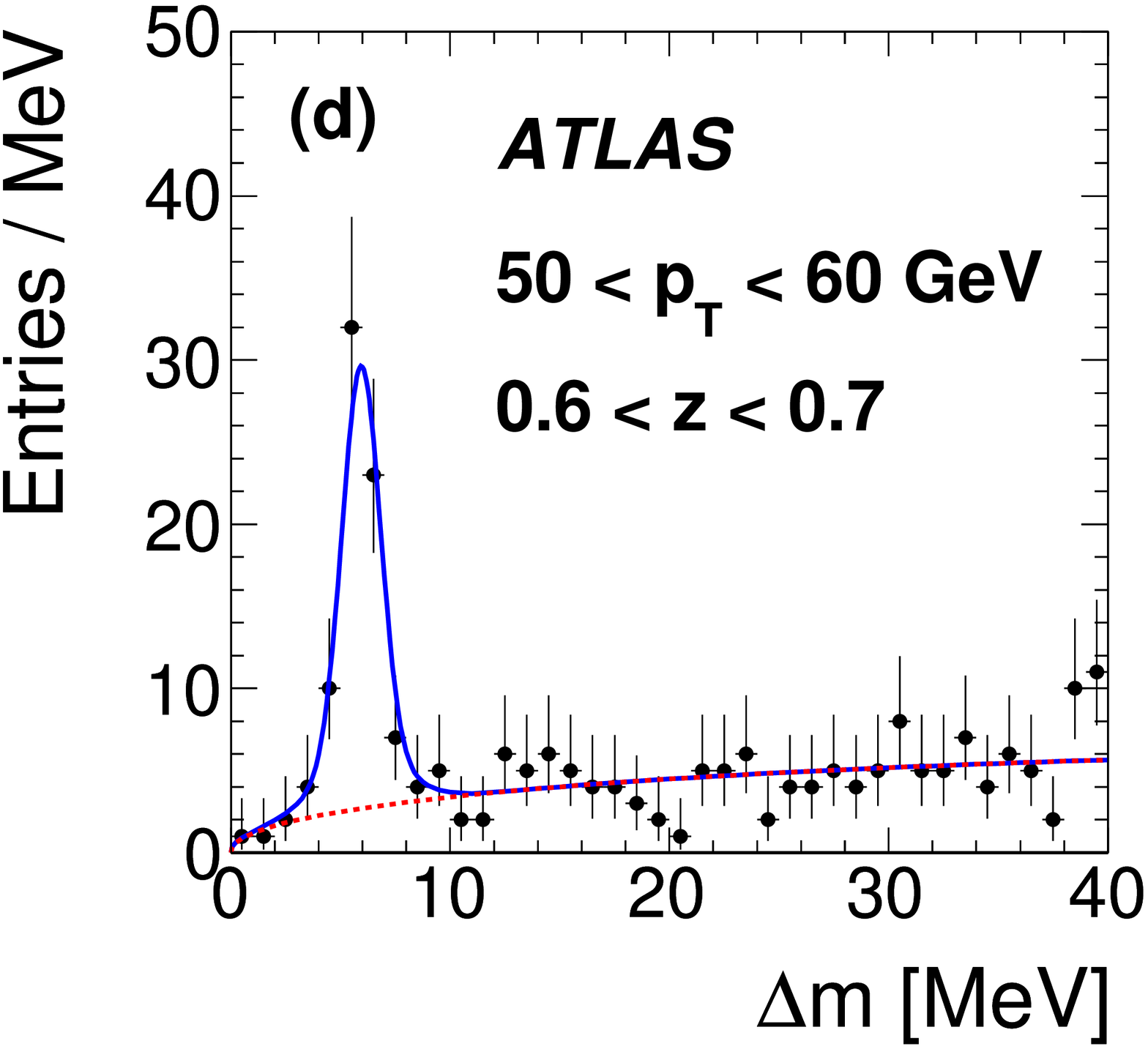}\\\vspace{1mm}
\includegraphics[width=0.235\textwidth]{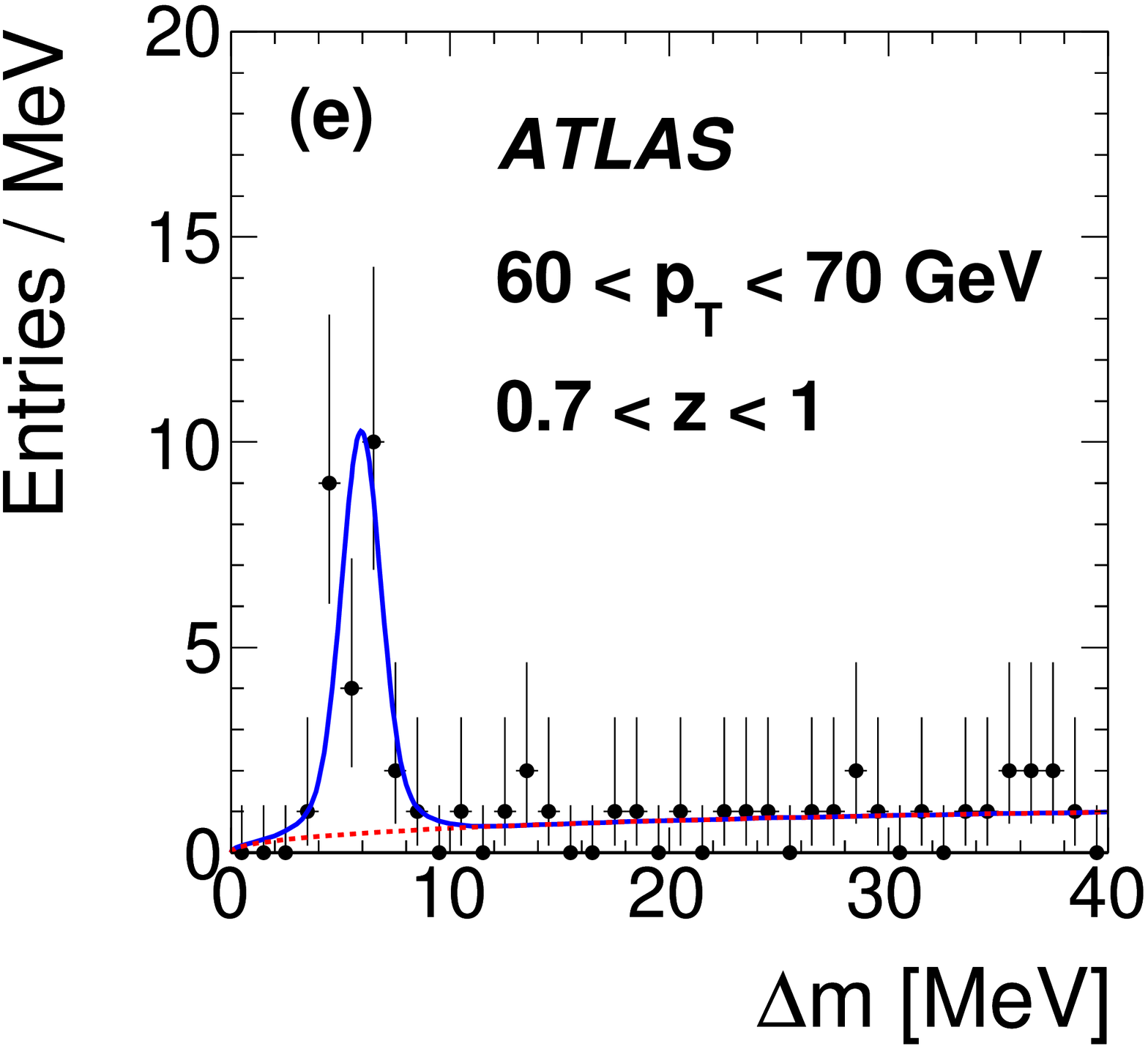}\hspace{1mm}
\includegraphics[width=0.235\textwidth]{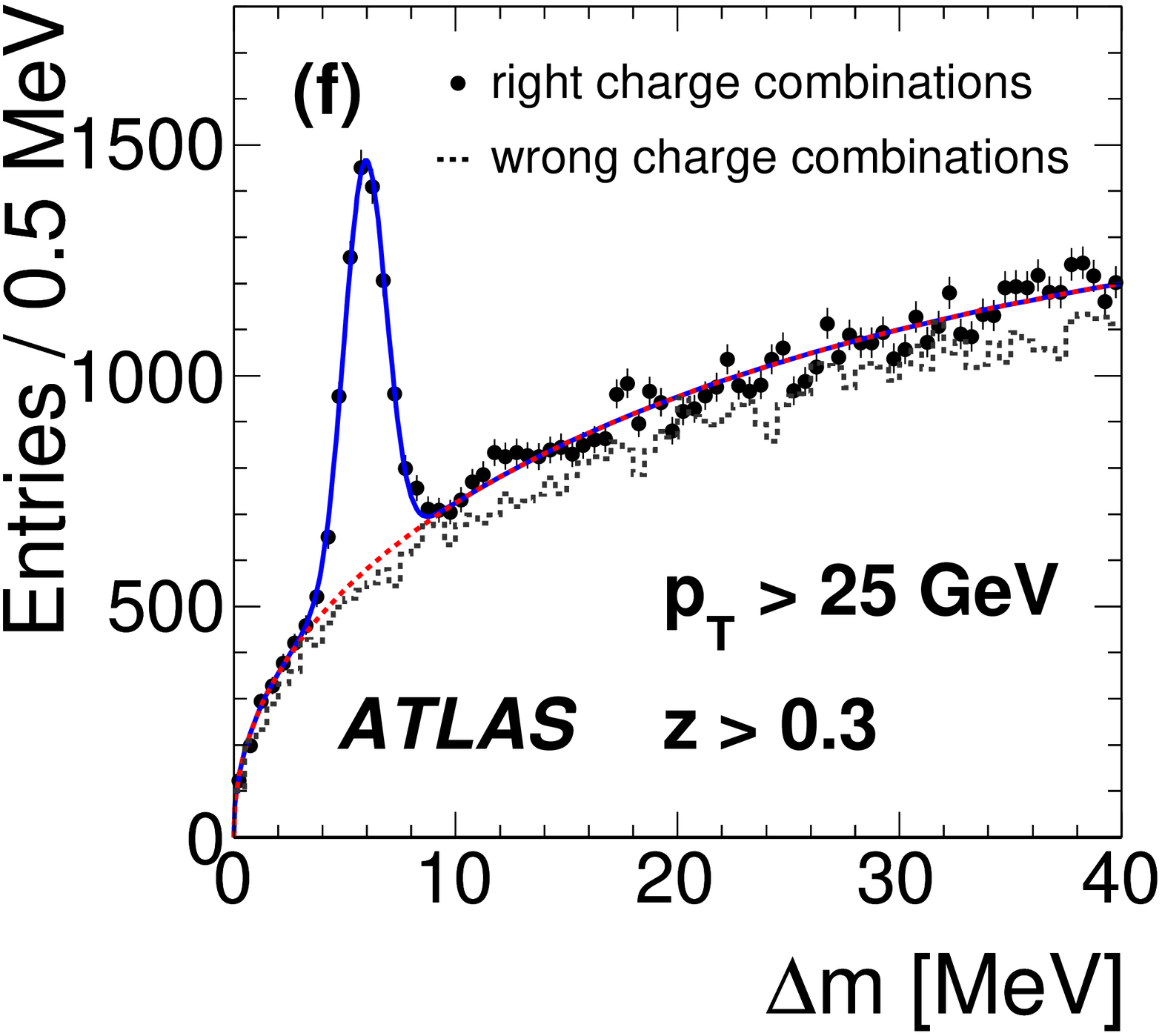}
\caption{Examples of the distributions of the mass difference of the
$D^{*+}\to D^0\pi^+$ and its charge conjugate inside jets for
different $p_{\rm T}$ and $z$ bins.
The solid line is the fit result. The dotted line
represents the background component. For comparison, 
the distribution of the difference between
the invariant mass of the wrong sign candidates,
$m(K^\pm\pi^\pm\pi^\mp)-m(K^\pm\pi^\pm)-m(\pi^\mp)$, is also
shown with a dashed line in Figure~(f),
where identical event selection criteria as the signal reconstruction have
been applied except that the two tracks from the $D^0$ ($\bar{D}^0$)
candidates are required to have the same charge.
No structure is observed and the distribution of wrong sign candidates can be
also well described by the same PDF as the signal candidates with slightly
different values of the PDF parameters.}
\label{fig:dmfit}
\end{center}
\end{figure}

The $D^{*\pm}$ jet yield is extracted from  the distribution of
$\Delta m=m(K^\mp\pi^\pm\pi^\pm)-m(K^\mp\pi^\pm)-m(\pi^\pm)$, where
$m(K^\mp\pi^\pm\pi^\pm)$ is the invariant mass of the $D^{*\pm}$ candidate
and $m(K^\mp\pi^\pm)$ is the invariant mass of the $D^0$ ($\bar{D}^0$)
candidate. The signal probability density function (PDF) is modeled as a
double Gaussian with equal mean based on MC studies,
and the background is characterized by $\Delta m^a\,\text{e}^{b\Delta m}$,
where $a$ and $b$ are free parameters in the
fit. The $D^{*\pm}$ jet candidate sample is divided into several bins in 
$p_{\rm T}$ and $z$ of the $D^{*\pm}$ jet. A simultaneous unbinned maximum
likelihood fit is then performed. In the fit, the parameters of the signal and
background PDFs are constrained to be the same in each $p_{\rm T}$ and $z$ bin,
and the fit returns the normalizations of the signal and the background.
The assumption of shapes being constant with $p_{\rm T}$ and $z$ has been checked in MC.
After applying all the event selection criteria, a total of $4282 \pm 93$
$D^{*\pm}$ jet signal candidates are obtained, where the error is statistical only.
Examples of the $\Delta m$ distribution for the data are shown in Fig.~\ref{fig:dmfit}
together with the fit result.

\section{Unfolding}

The signal yield of the reconstructed $D^{*\pm}$ jets is extracted in
bins of the jet $p_{\rm T}$ and $z$. If detector resolution effects are
negligible, the $D^{*\pm}$ jet production rate can be calculated as
\begin{equation}
{\cal R}(p_{\rm T},z)=\frac{N^{\rm reco}_{D^{*\pm}}(p_{\rm T},z)/\epsilon_{D^{*\pm}}(p_{\rm T},z)}
{{\cal B}(D^{*\pm}\to K^\mp\pi^\pm\pi^\pm)N^{\rm reco}_{\rm jet}(p_{\rm T})/\epsilon_{\rm jet}(p_{\rm T})},
\end{equation}
where $\cal B$  is the decay branching fraction, $\epsilon_{D^{*\pm}}$ is the
trigger and reconstruction efficiency of $D^{*\pm}$ jets identified by the decay
$D^{*\pm} \to K^{\mp}\pi^{\pm}\pi^{\pm}$, $\epsilon_{\rm jet}$ is the trigger
and reconstruction efficiency of inclusive jets, $N^{\rm reco}_{D^{*\pm}}$
and $N^{\rm reco}_{\rm jet}$ are the numbers of reconstructed $D^{*\pm}$ and
inclusive jets, respectively. MC studies show that for a given $D^{*\pm}$ jet
that passes the trigger selection and falls within the kinematic range of the
measurement, the probability to reconstruct it offline is between $15\%$ and $45\%$,
depending on the jet  $p_{\rm T}$ and $z$. However, the
reconstructed  values of the $p_{\rm T}$ and $z$ will be
different from the true values at the particle level due
to the finite resolution of the jet energy measurement.
In order to obtain the true distributions
at the particle level from the measured quantities, a Bayesian iterative
unfolding algorithm~\cite{D'Agostini:1994zf} is used to correct for the
detector efficiency and bin-to-bin migration due to the detector resolution.
The $D^{*\pm}$ jet production rate is subsequently calculated as
\begin{equation}
{\cal R}(p_{\rm T},z)=\frac{N_{D^{*\pm}}(p_{\rm T},z)}
{{\cal B}(D^{*\pm}\to K^\mp\pi^\pm\pi^\pm)N_{\rm jet}(p_{\rm T})},
\end{equation}
where $N_{D^{*\pm}}$ and $N_{\rm jet}$ are the number of the $D^{*\pm}$
and inclusive jets after unfolding, respectively.

\begin{figure}[!htb]
\begin{center}
\includegraphics[width=0.235\textwidth]{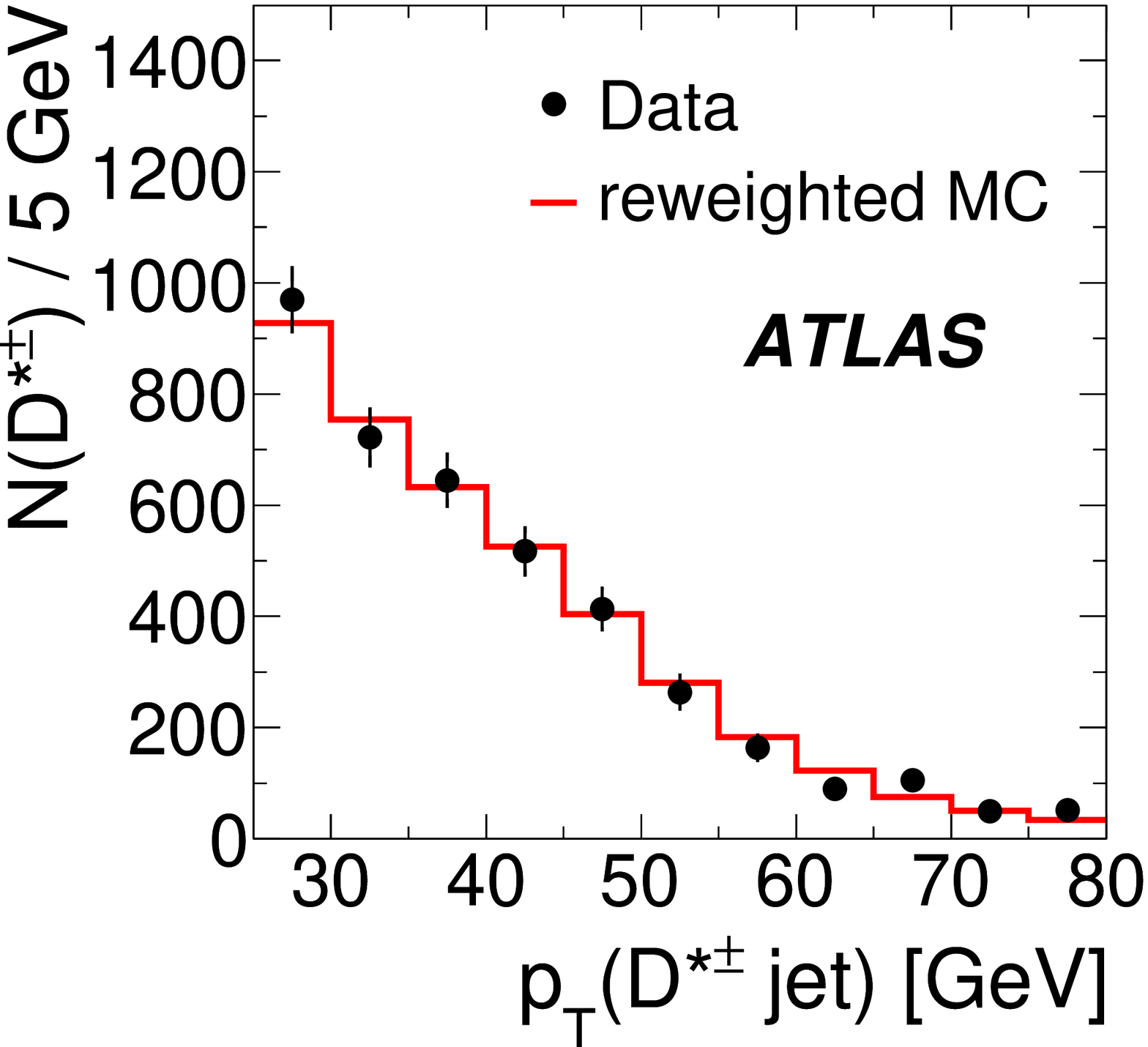}\hspace{1mm}
\includegraphics[width=0.235\textwidth]{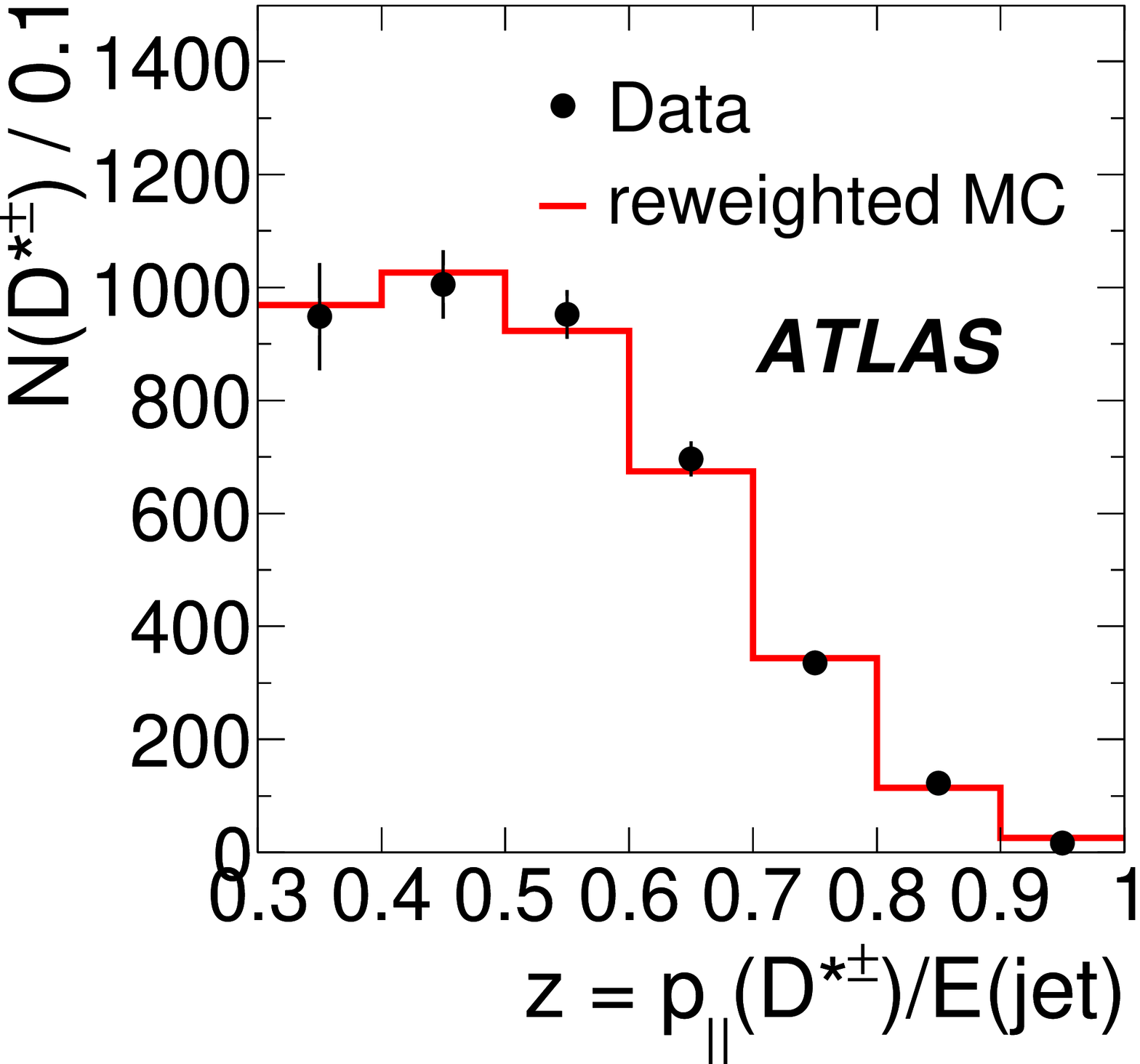}
\caption{Comparison of the $p_{\rm T}$ and $z$ distributions of the
$D^{*\pm}$ jets between data and MC. The $D^{*\pm}$ decay products
are matched to jets at the particle level by using a geometrical
matching with $\Delta R < 0.6$. The MC distributions are then reweighted
and subsequently normalized to have the same number of events as the data.}
\label{fig:reweighting}
\end{center}
\end{figure}

The corrections of this algorithm are based on a response matrix that is
derived from MC simulated events, which encapsulates the probability for
a true $D^{*\pm}$ jet at the particle level with a particular $p_{\rm T}$
and $z$ to be reconstructed in any possible $p_{\rm T}$ and $z$ bin. 
The MC $p_{\rm T}$ and $z$ distributions of $D^{*\pm}$ jets are reweighted
to match the measured distributions, the comparison is shown in
Fig.~\ref{fig:reweighting}. Similar correction procedures are also
applied to the reconstructed inclusive jets to obtain the number
of jets at the particle level as a function of the jet $p_{\rm T}$.

The unfolding algorithm has been validated using MC simulated events 
and no bias is observed. To evaluate the statistical uncertainties on the unfolded variables,
1000 ensembles of the unfolding sample are generated. For each ensemble, the
number of reconstructed jets in each bin is generated randomly according to
a Gaussian distribution, where its mean is the number of jets reconstructed
in that bin before unfolding, and its width is the corresponding statistical
uncertainty. The unfolding is subsequently performed for each ensemble. The deviations
of the numbers of jets after unfolding with respect to the nominal results
are fitted to a Gaussian distribution. The means of the Gaussian
distributions are found to be consistent with zero. The widths of those
Gaussian distributions are taken as the statistical uncertainties of the
measured numbers of jets after unfolding.

\section{Systematic Uncertainties}

\begin{figure*}[p]
\begin{center}
\vspace{15mm}
\includegraphics[width=0.49\textwidth]{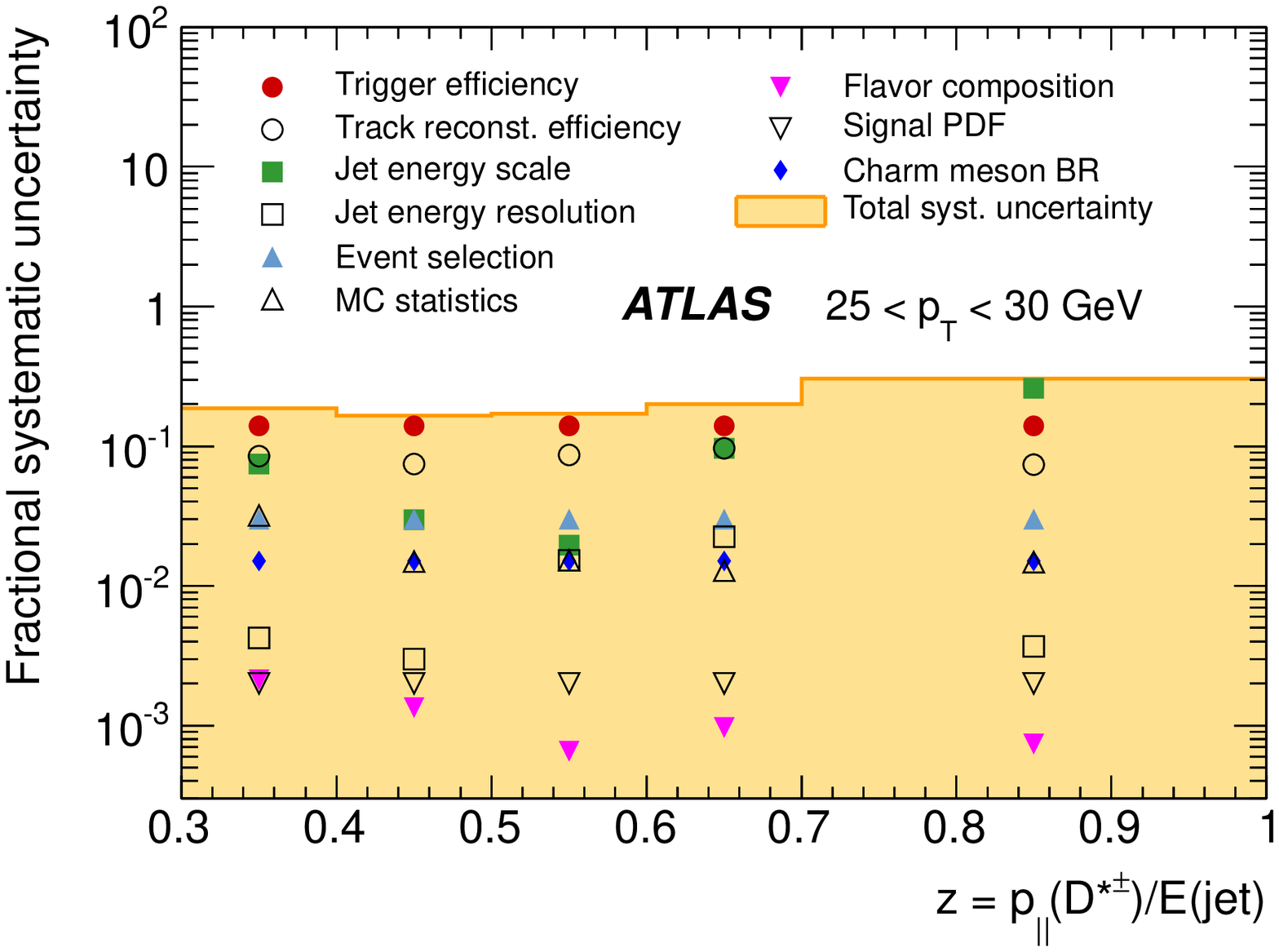}
\includegraphics[width=0.49\textwidth]{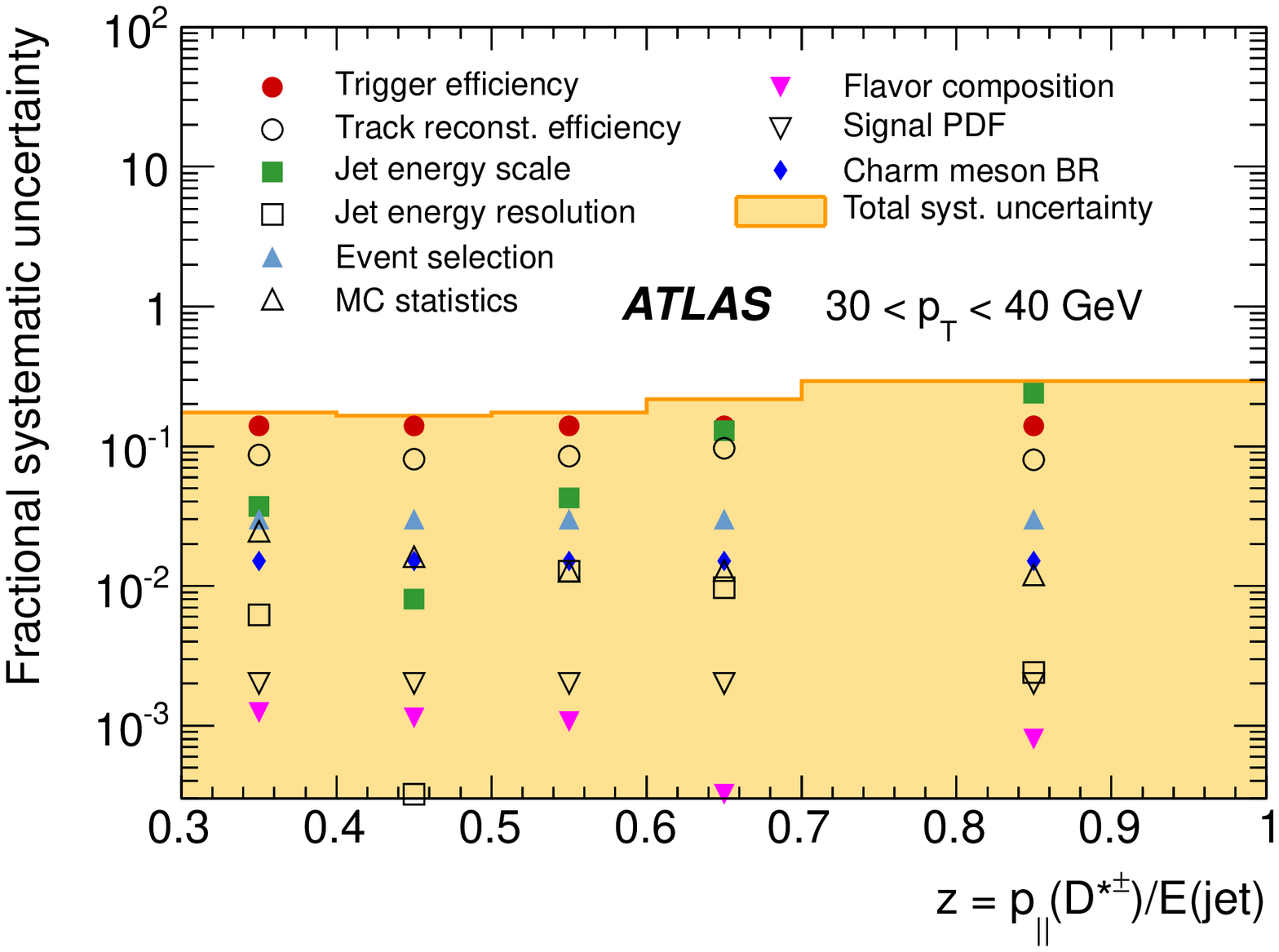}\\\vspace{2mm}
\includegraphics[width=0.49\textwidth]{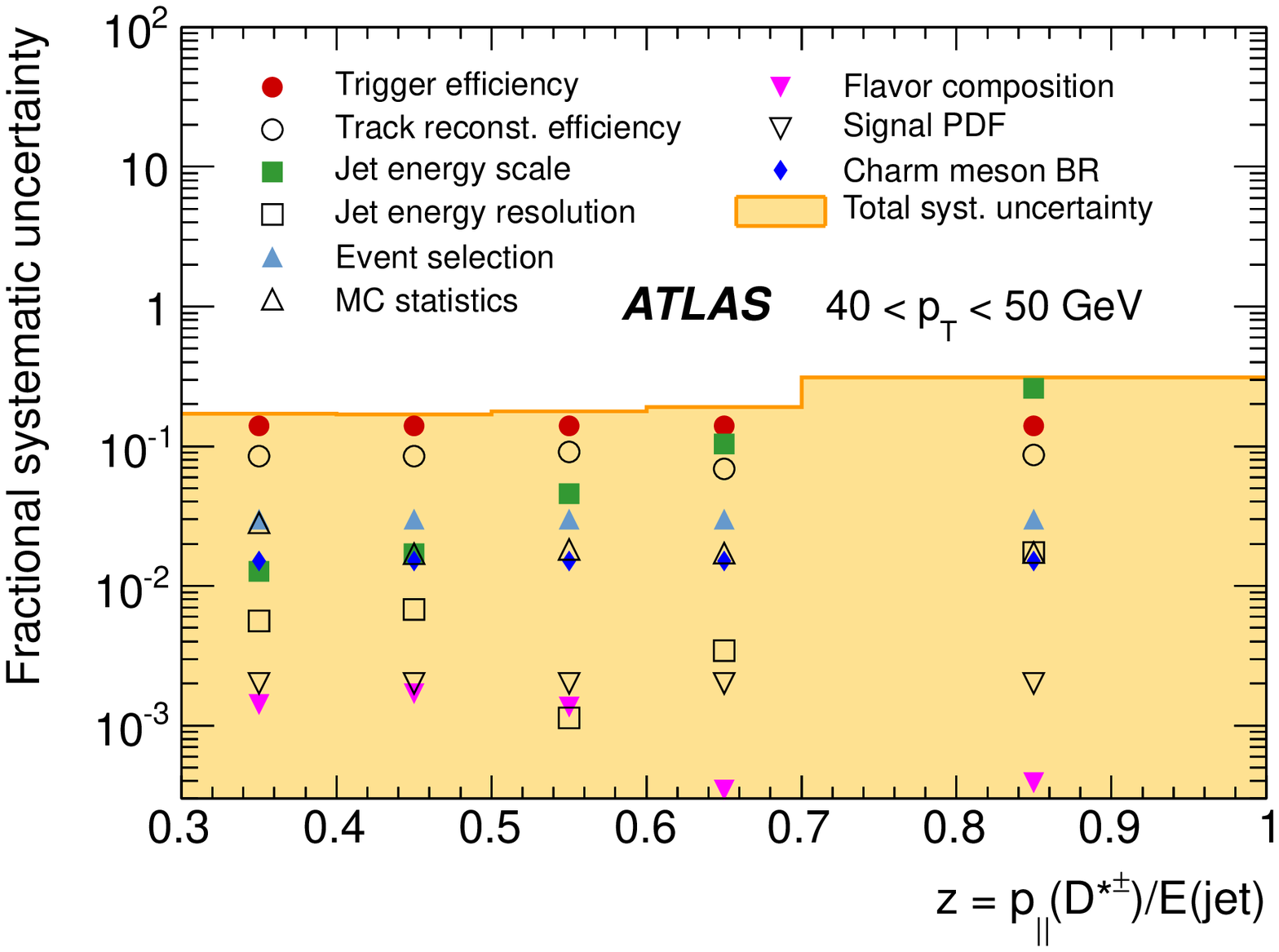}
\includegraphics[width=0.49\textwidth]{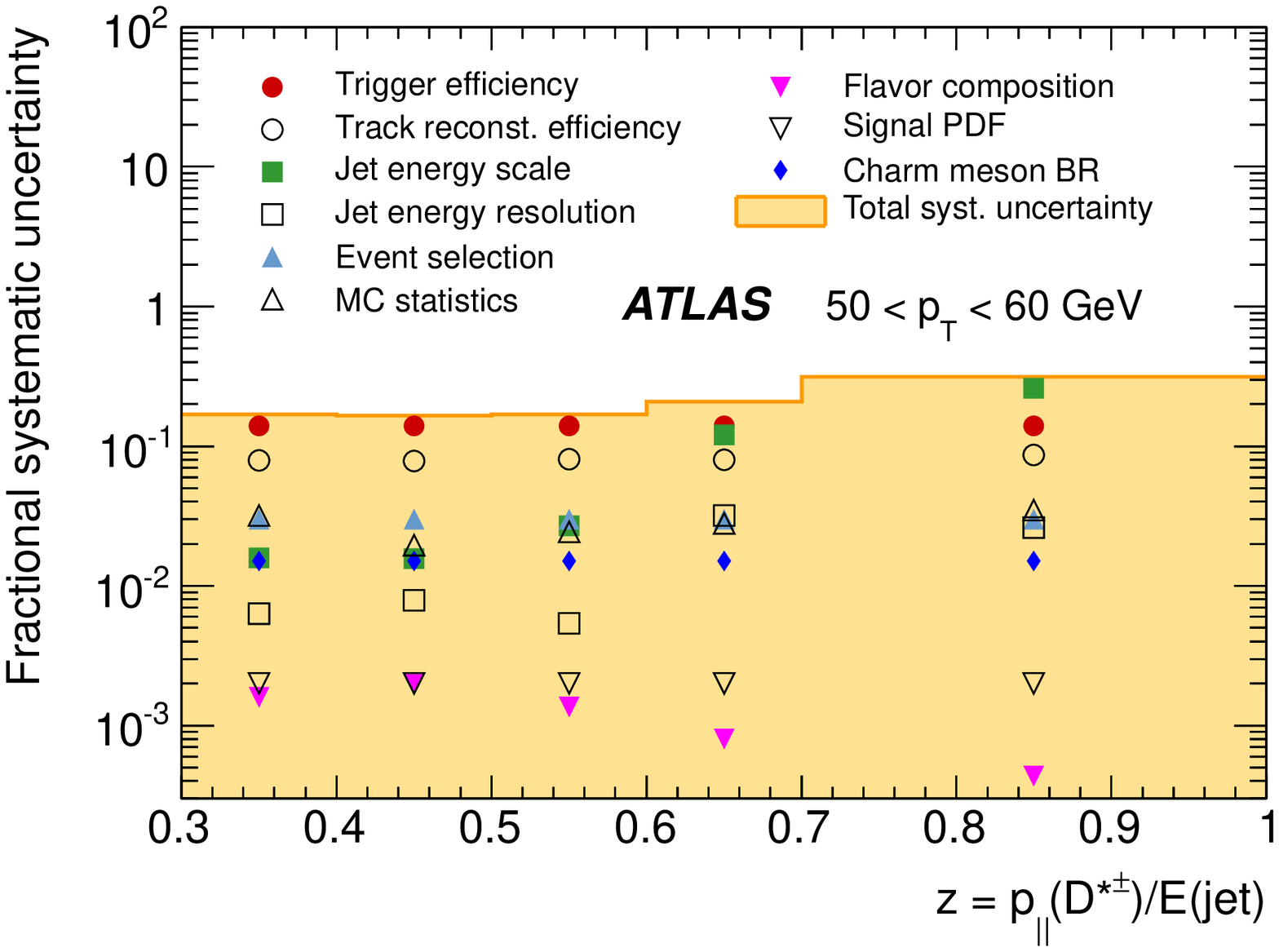}\\\vspace{2mm}
\includegraphics[width=0.49\textwidth]{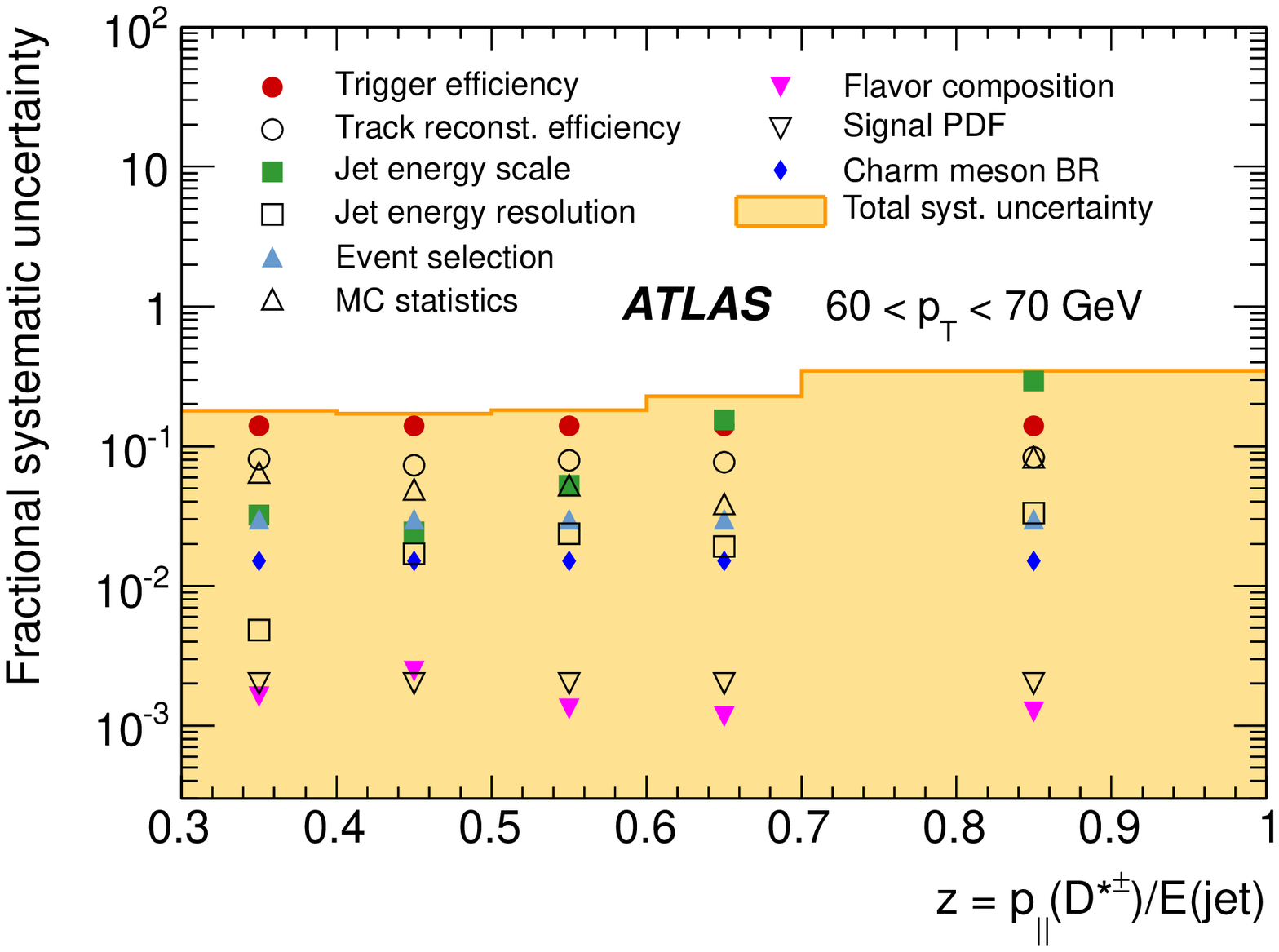}
\includegraphics[width=0.49\textwidth]{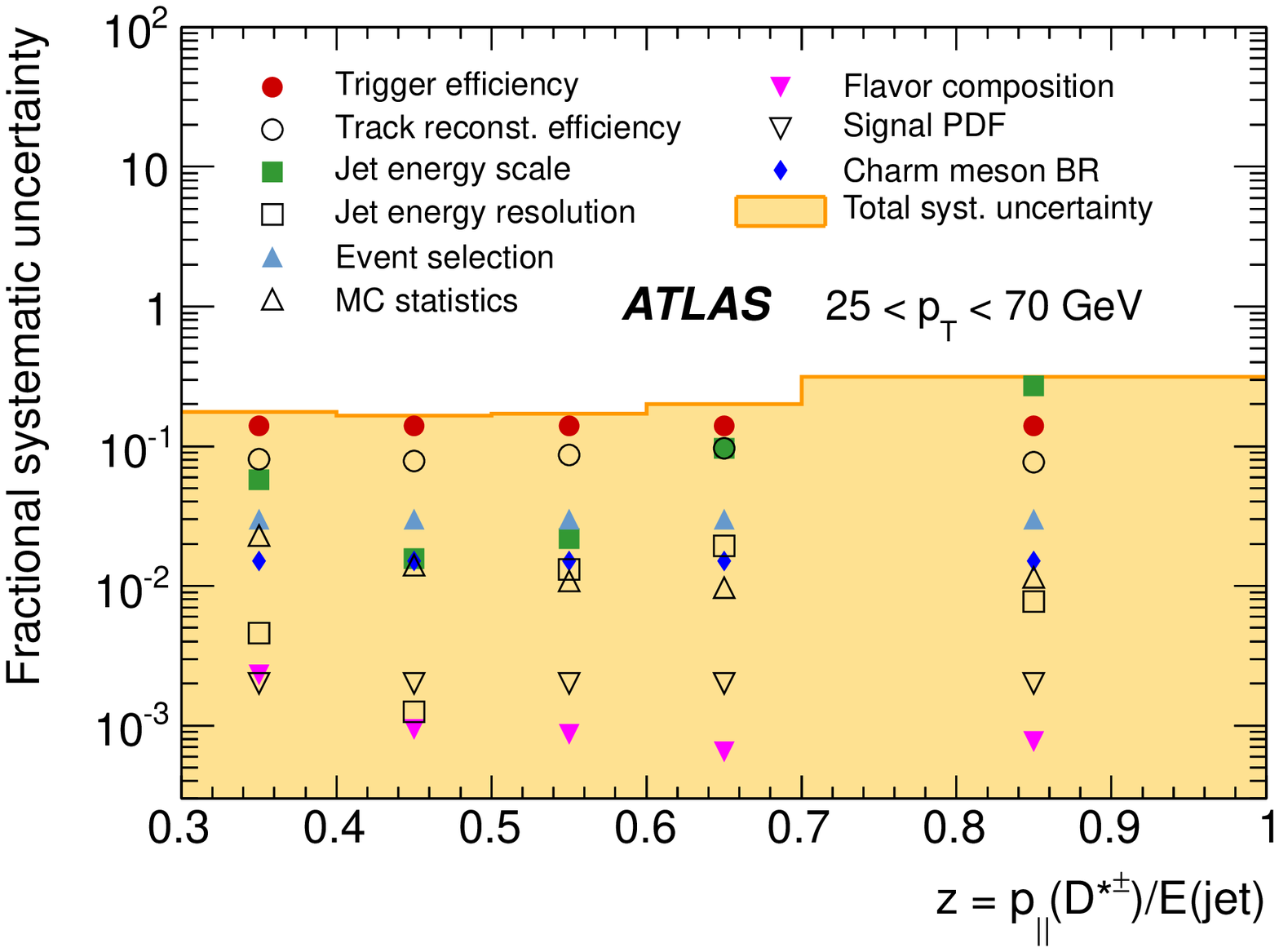}
\caption{Relative systematic uncertainties of the measured $D^{*\pm}$ jet
production rate ${\cal R}(p_{\rm T},z)$ in different jet $p_{\text T}$ and $z$ bins.
The values corresponding to the integrated $p_{\rm T}$ range are shown in the
bottom right Figure.}
\label{fig:syst}
\end{center}
\end{figure*}

The fractional systematic uncertainties of the measured $\cal R$ in each bin
of $p_{\rm T}$ and $z$ are shown in Fig.~\ref{fig:syst} while the total uncertainties
are summarized in Table~\ref{tab:R_pt_z_final}. A brief description of the sources
of systematic uncertainties in the measurement and how they are estimated
is given below.

Trigger efficiency effects largely cancel in the calculation of the 
ratio $\cal R$. However, the different flavor composition of $D^{*\pm}$ jets
and inclusive jets could cause differences in the trigger efficiencies of the
two samples. To study the effect of these possible differences on $\cal R$, a
variable  $r=N^{\rm reco}_{D^{*\pm}}/N^{\rm reco}_{\rm jet}$ is defined, where
$N^{\rm reco}_{D^{*\pm}}$ is the number of reconstructed $D^{*\pm}$ jets and
$N^{\rm reco}_{\rm jet}$ is the number of reconstructed inclusive jets. 
Subsequently a double ratio $\rho$ is defined as
$\rho={r_{\rm jet\; trig}}/{r_{\rm MBTS}}$, where
${r_{\rm jet\; trig}}$ and ${r_{\rm MBTS}}$ are the ratio $r$ measured in the
data collected by the L1 calorimeter jet trigger and the MBTS trigger, respectively. 
Since the MBTS trigger makes no jet selection in contrast
to the L1 jet triggers, the double ratio $\rho$ gives a good estimate of the
size of flavor-dependent trigger efficiency effects. The values of $\rho$ determined
in data and MC are found to agree within the statistical uncertainty.
Further comparisons between the values of the double ratio $\rho$ measured
in data and MC simulation are made as a function of the jet $p_{\rm T}$, $z$ and
$\eta$, and no significant difference between data and the MC
simulation is observed. As a result, the relative statistical uncertainty 
($14\%$) of the measured $\rho$ in the data is taken as the relative uncertainty
of the measurement due to potential bias from trigger effects. This large
systematic uncertainty is due to the limited size of the data sample
from the MBTS trigger, which was heavily prescaled during data taking.

To estimate the systematic uncertainties due to track reconstruction
efficiencies, which only affect the $D^{*\pm}$ jets, a weight
factor $w_{\rm trks}$ is assigned to each reconstructed $D^{*\pm}$ jet candidate in
the MC simulation:
\begin{equation}
w_{\rm trks}= (1 + s_{K})  \times (1 + s_{\pi_1}) \times (1 + s_{\pi_2}),
\end{equation}
where $s_{K}$, $s_{\pi_1}$ and $s_{\pi_2}$ are $\pm 1\sigma$
of the uncertainties on the
track reconstruction efficiencies for the $D^{*\pm}$ decay daughters. These uncertainties
are derived from data as a function of track $p_{\rm T}$ and $\eta$~\cite{:2010ir}.
The response matrix for $D^{*\pm}$ jets is recalculated using values of 
$w_{\rm trks}$ with the individual factors: $s_{K}$, $s_{\pi_1}$ and $s_{\pi_2}$, 
all positive or all negative and new values of $\cal R$ are derived.
 The deviations of the newly measured $D^{*\pm}$
production rates in each $p_{\rm T}$ and $z$ bin with respect to their nominal
measurement values are taken as the corresponding systematic uncertainties.
The overall relative systematic uncertainty on $\cal R$ integrated
over $p_{\rm T}$ and $z$ due to the track reconstruction efficiency is $8\%$.

The systematic uncertainty on the jet energy scale is evaluated in
Ref.~\cite{:2010wv}. The maximum JES uncertainty in the central
region is approximately $6.5\%$ for jets with $20<p_{\rm T}<70\gev$.
The uncertainties of jets in the end-cap regions are slightly larger
because additional uncertainties due to the intercalibration are added. 
This brings the uncertainties in the end-cap region up to approximately
$9\%$ for jets with $20<p_{\rm T}<70\gev$. To estimate the corresponding
systematic uncertainties of the measured $D^{*\pm}$ jet production rate,
the values of the reconstructed $p_{\rm T}$ and $z$ of the $D^{*\pm}$ jets
and $p_{\rm T}$ of the inclusive jets are varied coherently in the MC
according to the JES uncertainties when
calculating their response matrices. Measurements of the $D^{*\pm}$
jet production rates are then performed with the new response matrices.
The deviations of the measured $D^{*\pm}$ jet production rate in each
$p_{\rm T}$ and $z$ bin with respect to their nominal values
are taken as the corresponding systematic uncertainties, which gives a
relative systematic uncertainty of $3\%$ in the measurement of $\cal R$
integrated over $p_{\rm T}$ and $z$.

Using MC simulated events, the JES of $D^{*\pm}$ jets and inclusive
jets are found to be slightly different in the low $p_{\rm T}$ region but
are consistent with each other in the high $p_{\rm T}$ region.
Varying the JES by this full difference, the corresponding systematic
uncertainty on the measured $\cal R$ is estimated to be less than $1\%$.

The resolution of the jet energy measurement has been verified to
be in agreement within $14\%$ between data and MC simulation for jets
in the pseudorapidity range $|\eta|<2.8$ using control 
samples~\cite{ATLAS-CONF-2010-054}. To estimate the corresponding
effects on the measurement, the nominal jet energy resolution in the MC
simulation is artificially degraded
to account for this uncertainty
for both the $D^{*\pm}$ and inclusive jets. The $D^{*\pm}$ jet production rate
measurements are then repeated using the newly calculated response matrices.
The deviations of the measured $D^{*\pm}$ production rate in each $p_{\rm T}$
and $z$ bin with respect to their nominal values are taken as the
corresponding systematic uncertainties, and are found to be below $1\%$.

In the analysis, an event selection of $L_{xy}>0$ is applied. The $D^{*\pm}$ jets can be
directly produced in $pp$ interactions ($c$ jets) or from $b$-hadron
decays ($b$ jets). The efficiency of the $L_{xy}>0$ cut depends on the
 fractions of $c$ and $b$ jets since they have different $L_{xy}$
distributions. The relative efficiency of the requirement of $L_{xy}>0$ is estimated by
comparing the $D^{*\pm}$ jet signal yields with and without such a selection.
Its value in data is measured to be $0.87\pm0.03$, consistent
with the MC predicted value of $0.890 \pm 0.003$, where the uncertainties
are statistical only. As a result, a $3\%$ relative systematic
uncertainty on the measured $\cal R$ is assigned independent of the jet $p_{\rm T}$ and $z$.

The measurement depends on the decay branching fraction of
$D^{*\pm}$~\cite{Nakamura:2010zzi}. The uncertainties of the decay branching
fractions give a $1.5\%$ relative systematic uncertainty of the measured
$\cal R$, independently of the jet $p_{\rm T}$ and $\eta$.

Other systematic sources considered in the measurement include the finite
size of the MC sample, and the signal and background PDFs. All of them are found to
have negligible effects on the measurement ($\sim 1\%$). The total systematic
uncertainty is calculated by summing the individual systematic uncertainties
in quadrature. 

\section{Results and Discussion}

\begin{table*}[htb]
\begin{center}
\caption{$D^{*\pm}$ jet production rate $\cal R$ calculated in each $p_{\rm T}$
and $z$ bin, and in the full $p_{\rm T}$ range. The first uncertainty shown here
is statistical and the second is systematic.}
\begin{tabular}{lccccccccccccccc}
\hline\hline
\multicolumn{16}{c}{${\cal{R}}(p_{\rm T},z)$ [$10^{-2}$]} \\
\hline
Jet $p_{\rm T}$ [GeV] &&& $0.3<z<0.4$ &&& $0.4<z<0.5$ &&& $0.5<z<0.6$ &&& $0.6<z<0.7$ &&& $0.7<z<1.0$ \\
$25-30$ &&& $0.94\pm0.11\pm0.18$ &&& $0.67\pm0.05\pm0.11$ &&& $0.46\pm0.03\pm0.08$ &&& $0.31\pm0.01\pm0.06$ &&& $0.27\pm0.01\pm0.08$ \\
$30-40$ &&& $0.81\pm0.09\pm0.15$ &&& $0.62\pm0.04\pm0.10$ &&& $0.47\pm0.02\pm0.08$ &&& $0.31\pm0.01\pm0.07$ &&& $0.25\pm0.01\pm0.07$ \\
$40-50$ &&& $0.71\pm0.09\pm0.13$ &&& $0.59\pm0.04\pm0.10$ &&& $0.44\pm0.03\pm0.08$ &&& $0.29\pm0.02\pm0.06$ &&& $0.23\pm0.02\pm0.07$ \\
$50-60$ &&& $0.63\pm0.09\pm0.11$ &&& $0.51\pm0.05\pm0.09$ &&& $0.37\pm0.03\pm0.06$ &&& $0.25\pm0.02\pm0.05$ &&& $0.23\pm0.02\pm0.07$ \\
$60-70$ &&& $0.62\pm0.16\pm0.12$ &&& $0.41\pm0.08\pm0.07$ &&& $0.38\pm0.07\pm0.07$ &&& $0.26\pm0.05\pm0.06$ &&& $0.24\pm0.05\pm0.08$ \\
$25-70$ &&& $0.87\pm0.08\pm0.16$ &&& $0.64\pm0.03\pm0.11$ &&& $0.46\pm0.02\pm0.08$ &&& $0.31\pm0.01\pm0.06$ &&& $0.26\pm0.01\pm0.08$ \\
\hline\hline
\end{tabular}
\label{tab:R_pt_z_final}
\end{center}
\end{table*}

\begin{figure*}[htb]
\begin{center}
\includegraphics[width=0.45\textwidth]{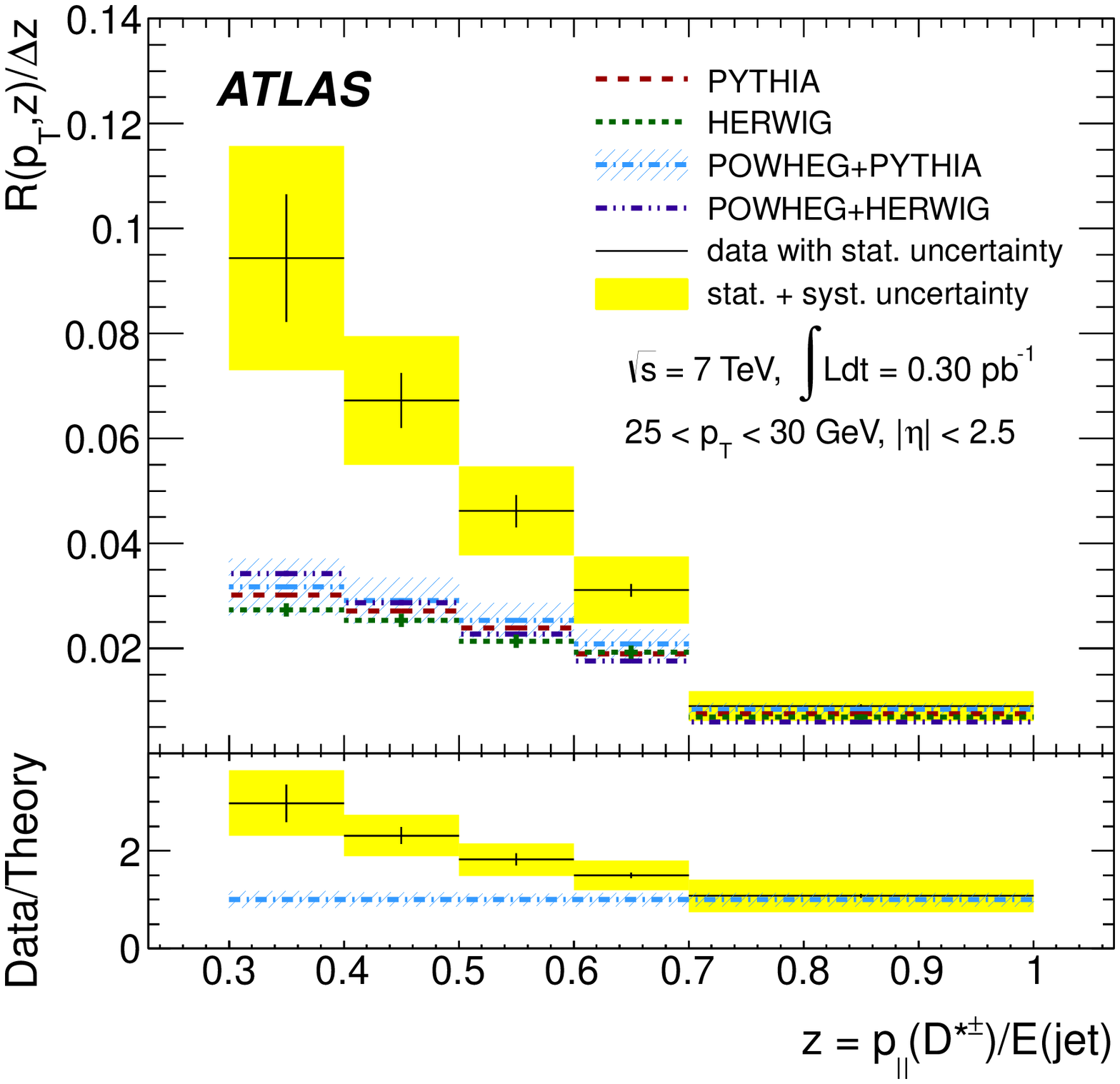}
\includegraphics[width=0.45\textwidth]{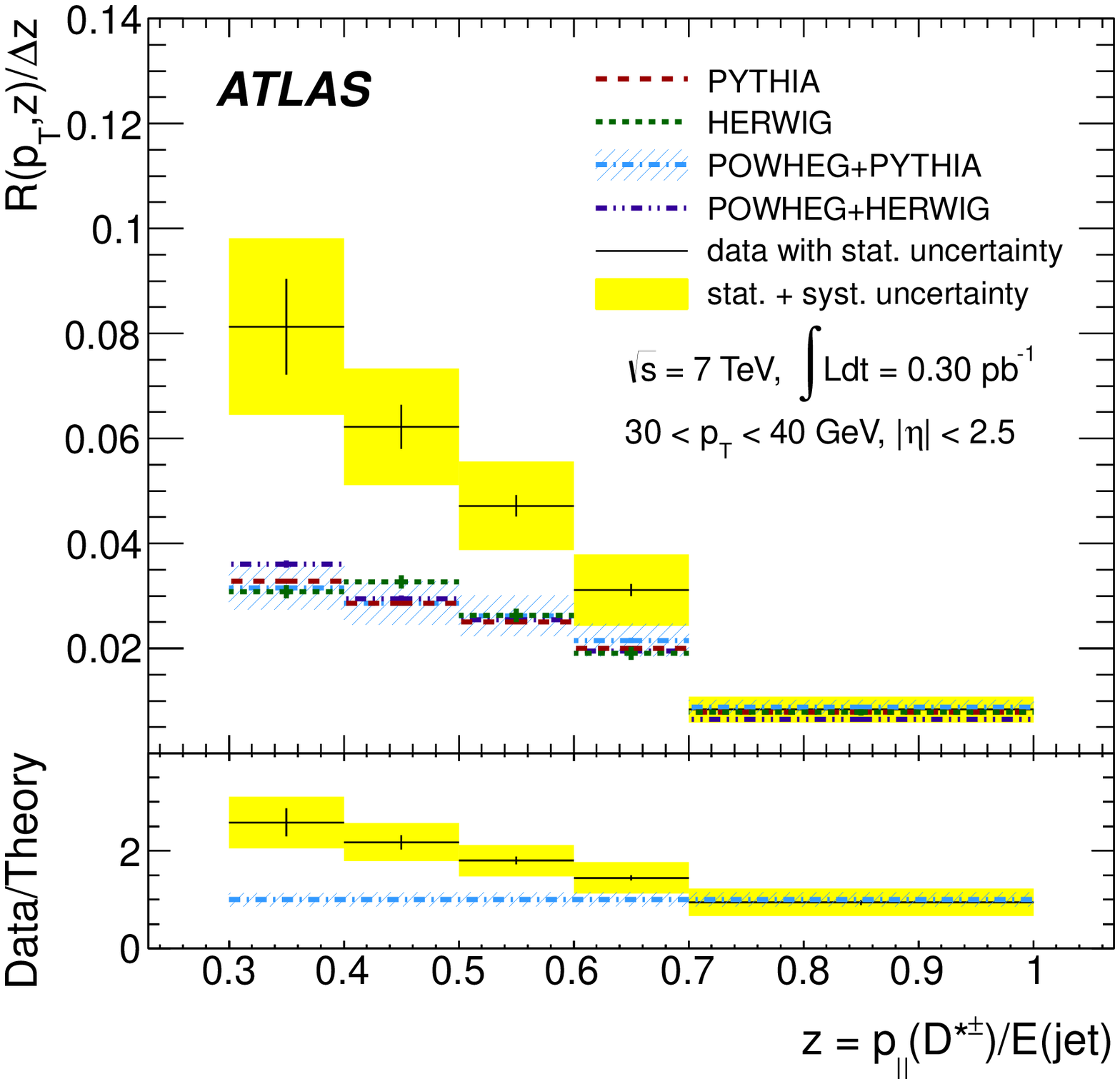}
\includegraphics[width=0.45\textwidth]{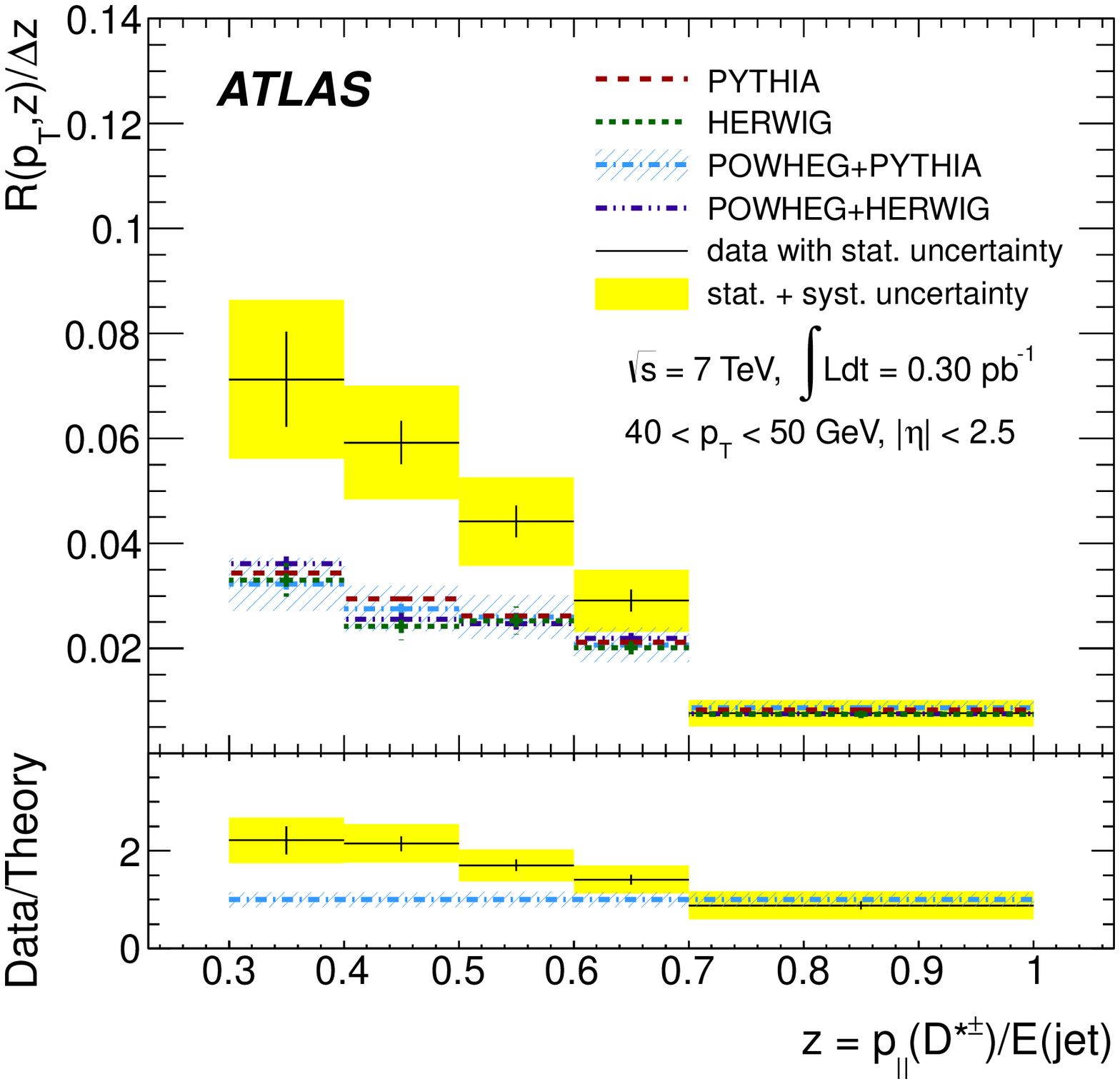}
\includegraphics[width=0.45\textwidth]{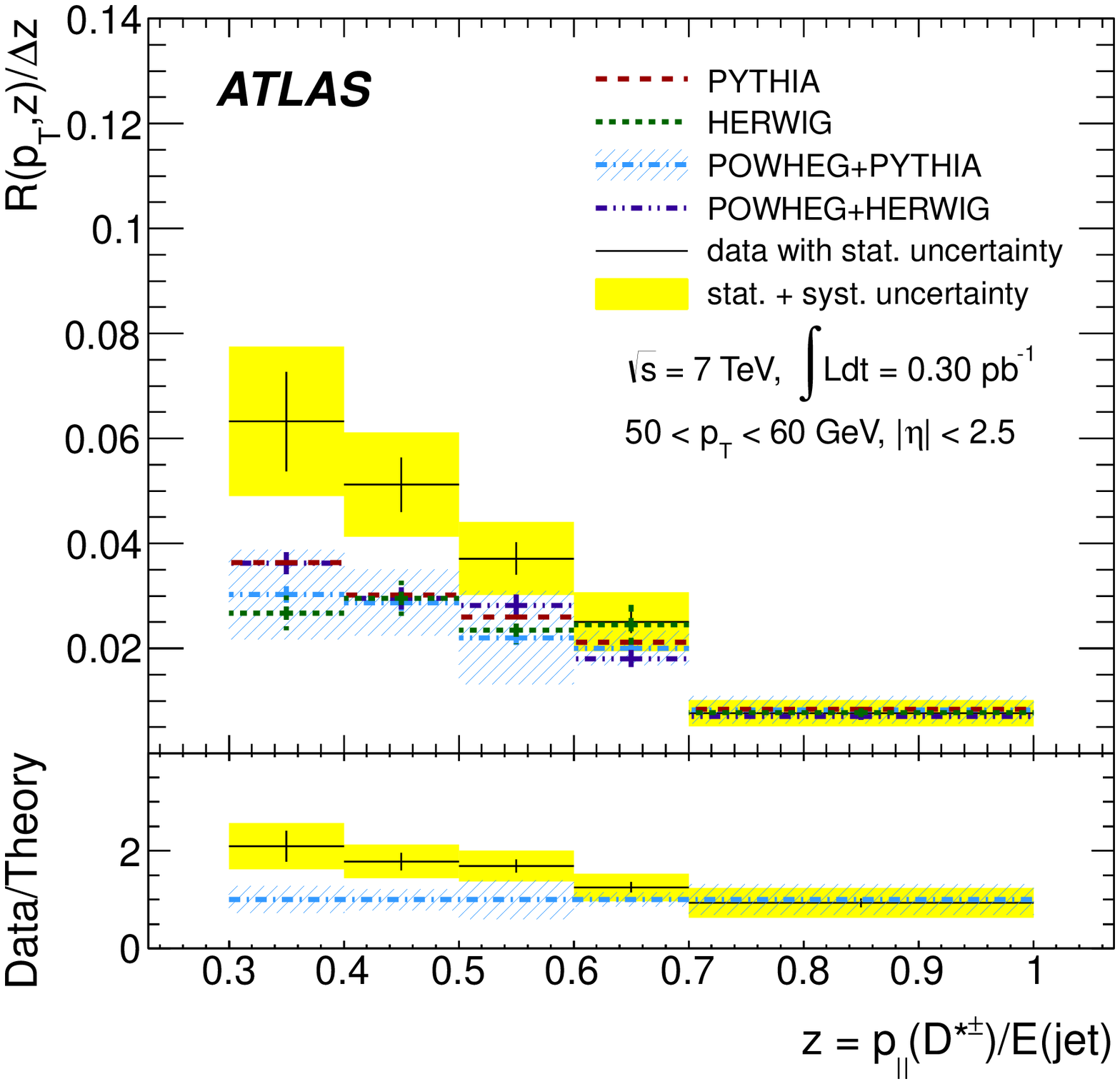}
\includegraphics[width=0.45\textwidth]{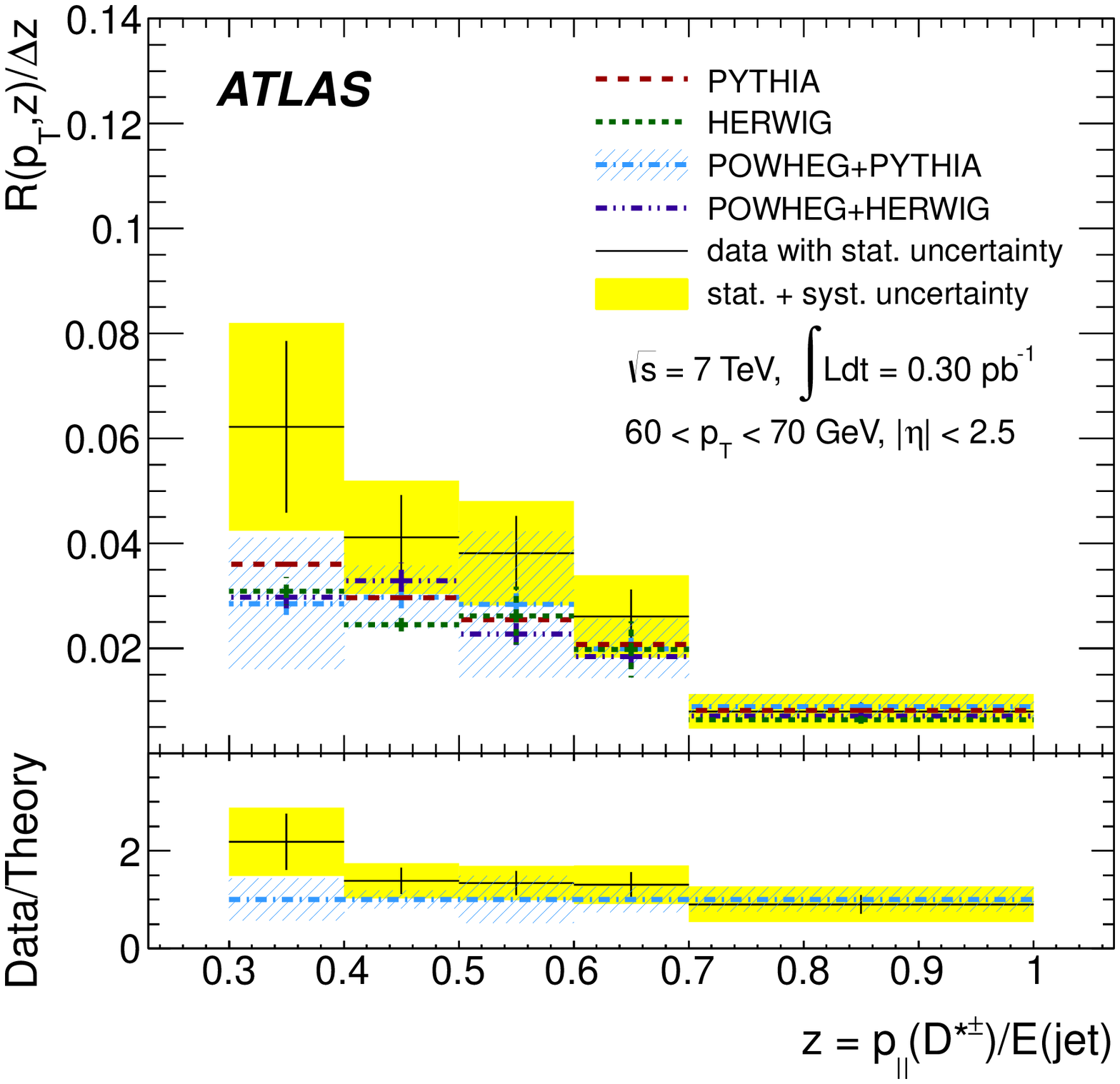}
\includegraphics[width=0.45\textwidth]{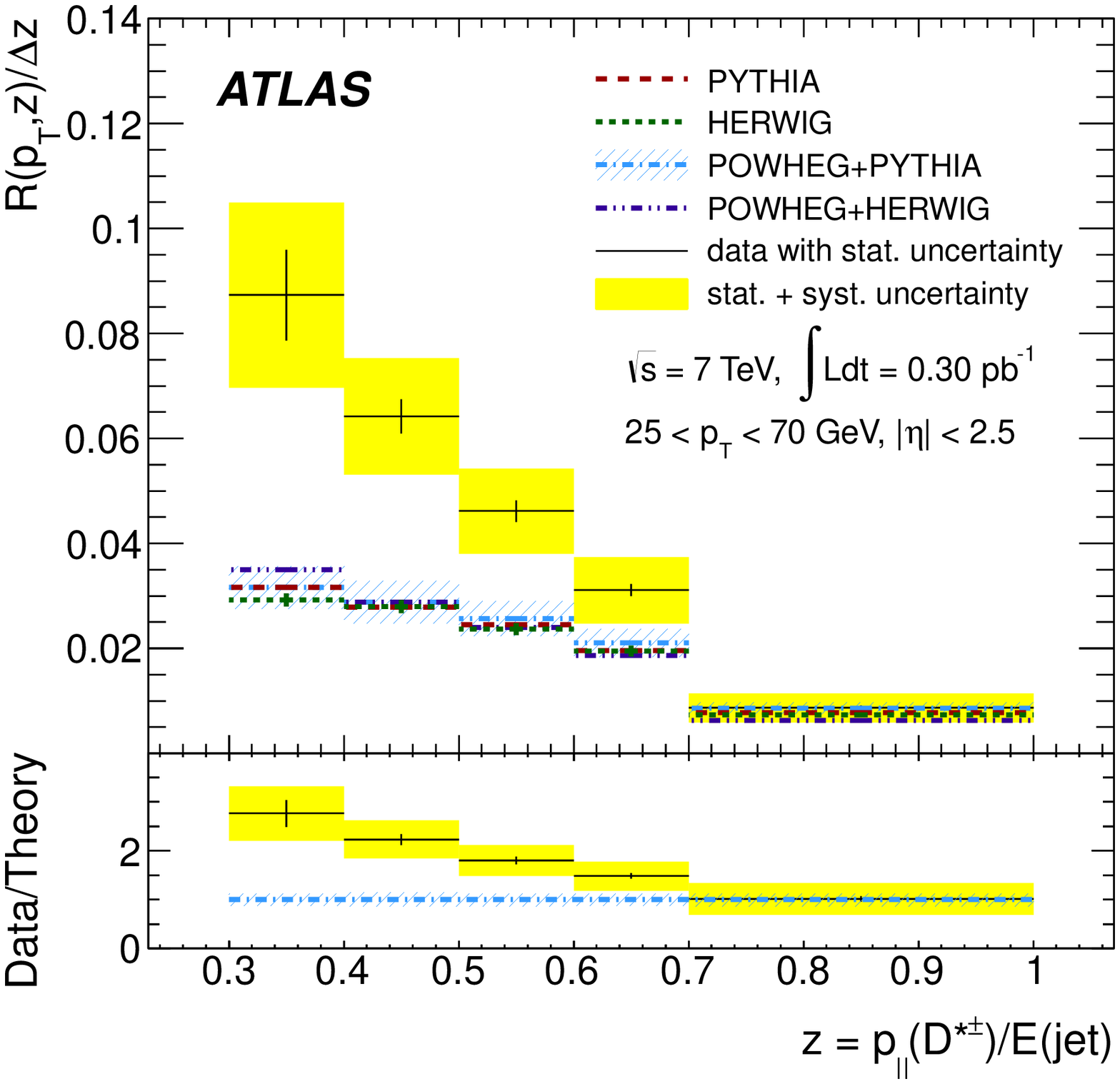}
\caption{Comparison of the $D^{*\pm}$ production rate
${\cal R}(p_{\rm T},z)/\Delta z$ in different jet $p_{\text T}$ and $z$ bins 
between the measurement and the MC predictions of {\footnotesize{PYTHIA}},
{\footnotesize{HERWIG}}, {\footnotesize{POWHEG}}+{\footnotesize{PYTHIA}}
and {\footnotesize{POWHEG}}+{\footnotesize{HERWIG}}.
The values corresponding to the integrated $p_{\rm T}$ range
are shown in the bottom right Figure.
The insets show the ratio of the measurement to the
{\footnotesize{POWHEG}}+{\footnotesize{PYTHIA}} prediction.}
\label{fig:theory_comp}
\end{center}
\end{figure*}

The measured $D^{*\pm}$ jet production rates $\cal R$ in each bin of $p_{\rm T}$
and $z$ are listed in Table~\ref{tab:R_pt_z_final} and shown in
Fig.~\ref{fig:theory_comp}. Integrating over all the $p_{\rm T}$
and $z$ bins, the production rate is found to be
\begin{equation}
{\cal R}=0.025\pm 0.001\stat \pm 0.004\syst,
\end{equation}
for $D^{*\pm}$ jets with transverse momentum between $25$ and $70\gev$, in
the range $|\eta|<2.5$, and with momentum fraction $0.3<z<1$. 

\begin{figure*}[htb]
\begin{center}
\includegraphics[width=0.47\textwidth]{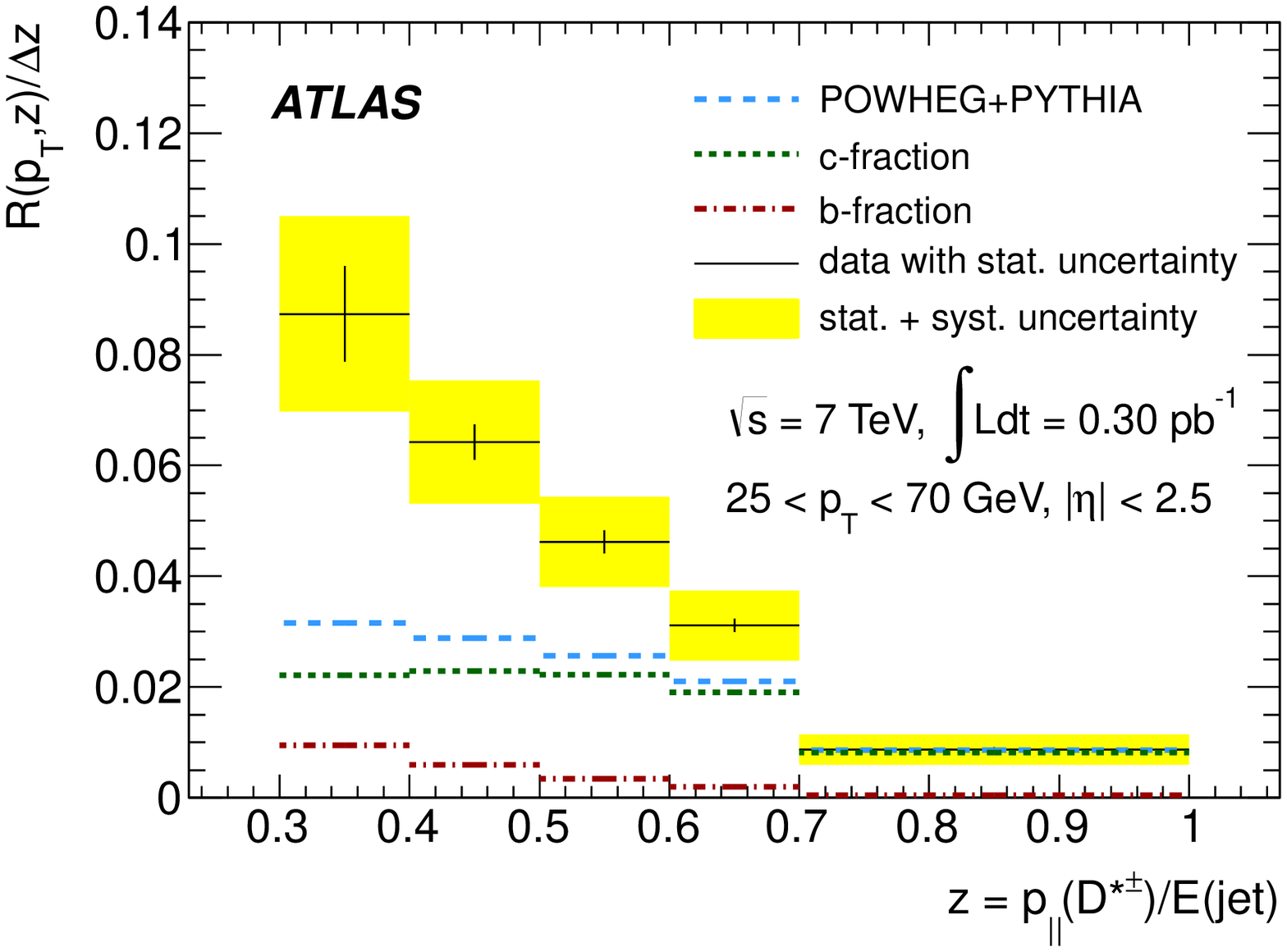}
\includegraphics[width=0.47\textwidth]{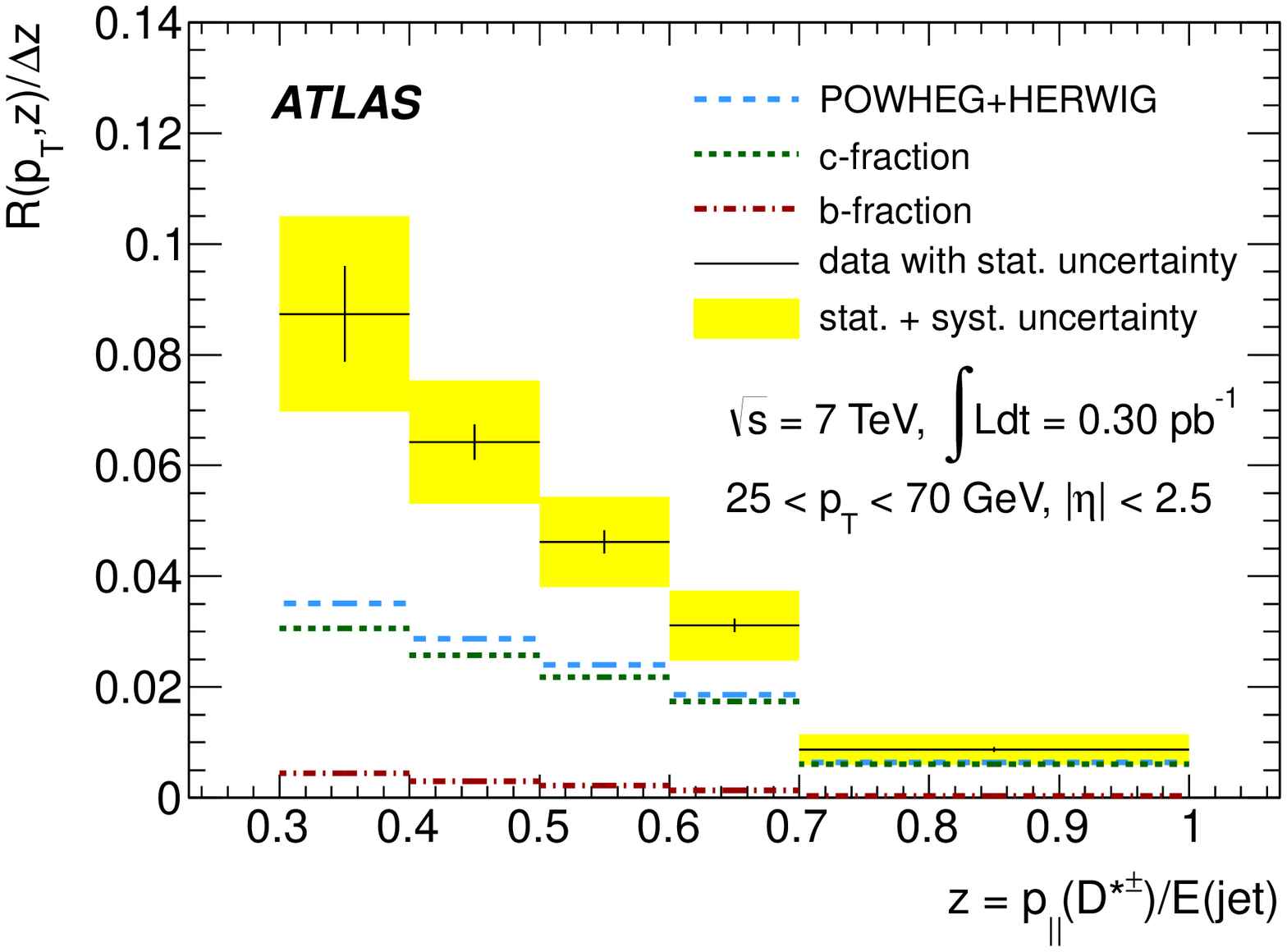}
\caption{Comparison of the $D^{*\pm}$ production rate
${\cal R}(p_{\rm T},z)/\Delta z$ in different jet $p_{\text T}$ and $z$ bins
between the measurement and the MC predictions of
{\footnotesize{POWHEG}}+{\footnotesize{PYTHIA}} (left) and
{\footnotesize{POWHEG}}+{\footnotesize{HERWIG}} (right).
The $c$- and $b$ fractions predicted by MC are also shown.
Only the central values of the MC predictions are shown here.}
\label{fig:cb_fraction}
\end{center}
\end{figure*}

Comparisons between the measurement and predictions from various MC
calculations are shown in Fig.~\ref{fig:theory_comp} as a function of
$z$ for different $p_{\rm T}$ ranges. The corresponding $c$ and $b$ jet
fractions predicted by MC are also shown in Fig.~\ref{fig:cb_fraction}.
The predicted values of $\cal R$ by {\footnotesize{PYTHIA}} and
{\footnotesize{POWHEG}}+{\footnotesize{PYTHIA}} are very similar, which is
also the case when comparing calculations from {\footnotesize{HERWIG}} and
{\footnotesize{POWHEG}}+{\footnotesize{HERWIG}}, as expected.
Since $\cal R$ is defined as the ratio between the number of $D^{*\pm}$ jets and
inclusive jets, the changes of total jet cross sections and $p_{\rm T}$
distributions between LO and NLO QCD calculations largely cancel.
The values of $\cal R$ predicted by MC calculations are lower than the data
by a factor 2 to 3 in the bins with lowest $z$, and this is especially
significant at low $p_{\rm T}$. The predictions are consistent with the
data for $z>0.7$ at all $p_{\rm T}$.
Integrating over all the $p_{\rm T}$ and $z$ bins, the
production rate $\cal R$ is estimated to be $0.0133 \pm 0.0008$ by
{\footnotesize{POWHEG}}+{\footnotesize{PYTHIA}}, which is just about half of the
measured value.

The various MC predictions share the feature that the $z$ distribution shape
is essentially independent of $p_{\rm T}$. In the data, there is a general
trend that ${\cal R}(p_{\rm T},z)$ falls with $p_{\rm T}$ for a fixed $z$ bin,
as is visible in Table~\ref{tab:R_pt_z_final}, in qualitative disagreement
with the MC prediction.

To further understand the discrepancies between the measurement and the
MC predictions, studies of the effects of various sources of systematic
uncertainty in the MC predictions are carried out using the
{\footnotesize{POWHEG}}+{\footnotesize{PYTHIA}} MC
program. The uncertainties on the calculated $\cal R$ are evaluated by varying
independently the renormalization and factorization scales between 0.5 and
2 times the default scale. The largest shift of $\cal R$ with respect to the
default calculation is taken as the corresponding systematic error due to
the uncertainties of the renormalization and factorization scales. Similarly
the possible systematic uncertainties associated with the charm and
bottom quark masses are estimated by varying them independently within $\pm0.2\,\gev$ and
$\pm0.25\,\gev$, respectively. 
The systematic uncertainties due to $f(c\to D^{*+})$ and $f(b\to D^{*+}X)$ are
evaluated by changing their values according to their measured
uncertainties~\cite{CLEO_1988,ARGUS,BABAR,CLEO_2004,BELLE,OPAL}.
Contributions from other sources, such as the
value of the strong coupling constant and the uncertainty of the parton
density function of the proton are much smaller and they are not taken into
account. The total systematic uncertainty of the MC calculations is computed
by summing each individual systematic uncertainty in quadrature. As shown in
Fig.~\ref{fig:theory_comp}, although the predicted values of $\cal R$ have
sizable systematic uncertainties in each bin, especially for
large $p_{\rm T}$ and $z$, the systematic errors become much smaller (less
than $10\%$) when integrating over the $p_{\rm T}$ bins.
The change of the calculated $\cal R$ in each bin is
dominated by the variation of $p_{\rm T}$ distributions of the $D^{*\pm}$ jets
and of the inclusive jets when different heavy quark masses and 
renormalization and factorization scales are used. Nevertheless, it is clear that
the systematic uncertainties that are considered in the MC calculation of $\cal R$ 
do not explain the discrepancies between data and MC predictions.

\section{Conclusions}

This paper reports a first measurement of $D^{*\pm}$ meson production
in jets in proton-proton collisions at a center-of-mass energy of
$\sqrt{s}=7\tev$ at the LHC using $0.30~\ipb$ of ATLAS data. The
production rate is found to be
$N(D^{*\pm})/N({\rm jet})=0.025\pm 0.001\stat \pm 0.004\syst$ for jets
with transverse momentum between $25$ and $70\gev$ in the range
$|\eta|<2.5$, and with $D^{*\pm}$ momentum fraction $0.3<z<1$.
Large discrepancies are observed between data and MC predictions for low $z$,
decreasing a little at higher $p_{\rm T}$. The $D^{*\pm}$ $z$ distributions
in data differ from the predictions of all the generators considered,
{\footnotesize{PYTHIA}}, {\footnotesize{HERWIG}} and {\footnotesize{POWHEG}},
both in overall normalization and shape.
The shapes of the $z$-distributions arising from $c$ and $b$ jets are expected to
be different. However, the differences observed between the data and MC predictions
cannot be explained by varying the mixture of $c$ and $b$ jets in the MC.
Contrary to the MC predictions, the measured $\cal R$ values listed in
Table~\ref{tab:R_pt_z_final} show a small, though monotonic decrease as a
function of the jet $p_{\rm T}$ in all the $z$ bins.
These observations indicate that the production of $c$ jets ($b$ jets) or their
fragmentation into $D^{*\pm}$ mesons is not well modeled in current MC generators.
These results show the need of further QCD refinements to improve the description
of high transverse momentum $D$-meson production in this new energy range of
hadron collisions.

\vspace{1mm}

\section{Acknowledgments}

We thank CERN for the very successful operation of the LHC, as well
as the support staff from our institutions without whom ATLAS could
not be operated efficiently.

We acknowledge the support of ANPCyT, Argentina; YerPhI, Armenia; ARC,
Australia; BMWF, Austria; ANAS, Azerbaijan; SSTC, Belarus; CNPq and FAPESP,
Brazil; NSERC, NRC and CFI, Canada; CERN; CONICYT, Chile; CAS, MOST and
NSFC, China; COLCIENCIAS, Colombia; MSMT CR, MPO CR and VSC CR, Czech
Republic; DNRF, DNSRC and Lundbeck Foundation, Denmark; ARTEMIS, European
Union; IN2P3-CNRS, CEA-DSM/IRFU, France; GNAS, Georgia; BMBF, DFG, HGF, MPG
and AvH Foundation, Germany; GSRT, Greece; ISF, MINERVA, GIF, DIP and
Benoziyo Center, Israel; INFN, Italy; MEXT and JSPS, Japan; CNRST, Morocco;
FOM and NWO, Netherlands; RCN, Norway; MNiSW, Poland; GRICES and FCT,
Portugal; MERYS (MECTS), Romania; MES of Russia and ROSATOM, Russian
Federation; JINR; MSTD, Serbia; MSSR, Slovakia; ARRS and MVZT, Slovenia;
DST/NRF, South Africa; MICINN, Spain; SRC and Wallenberg Foundation,
Sweden; SER, SNSF and Cantons of Bern and Geneva, Switzerland; NSC, Taiwan;
TAEK, Turkey; STFC, the Royal Society and Leverhulme Trust, United Kingdom;
DOE and NSF, United States of America.

The crucial computing support from all WLCG partners is acknowledged
gratefully, in particular from CERN and the ATLAS Tier-1 facilities at
TRIUMF (Canada), NDGF (Denmark, Norway, Sweden), CC-IN2P3 (France),
KIT/GridKA (Germany), INFN-CNAF (Italy), NL-T1 (Netherlands), PIC (Spain),
ASGC (Taiwan), RAL (UK) and BNL (USA) and in the Tier-2 facilities
worldwide.

\onecolumngrid
\clearpage

\begin{flushleft}
{\Large The ATLAS Collaboration}

\bigskip

G.~Aad$^{\rm 48}$,
B.~Abbott$^{\rm 111}$,
J.~Abdallah$^{\rm 11}$,
A.A.~Abdelalim$^{\rm 49}$,
A.~Abdesselam$^{\rm 118}$,
O.~Abdinov$^{\rm 10}$,
B.~Abi$^{\rm 112}$,
M.~Abolins$^{\rm 88}$,
H.~Abramowicz$^{\rm 153}$,
H.~Abreu$^{\rm 115}$
% ...
\bigskip

% ...
$^{10}$ Institute of Physics, Azerbaijan Academy of Sciences, Baku, Azerbaijan\\
$^{11}$ Institut de F\'isica d'Altes Energies and Departament de F\'isica de la Universitat Aut\`onoma  de Barcelona and ICREA, Barcelona, Spain\\
% ...
$^{48}$ Fakult\"{a}t f\"{u}r Mathematik und Physik, Albert-Ludwigs-Universit\"{a}t, Freiburg i.Br., Germany\\
$^{49}$ Section de Physique, Universit\'e de Gen\`eve, Geneva, Switzerland\\
% ...
$^{88}$ Department of Physics and Astronomy, Michigan State University, East Lansing MI, United States of America\\
% ...
$^{111}$ Homer L. Dodge Department of Physics and Astronomy, University of Oklahoma, Norman OK, United States of America\\
$^{112}$ Department of Physics, Oklahoma State University, Stillwater OK, United States of America\\
% ...
$^{115}$ LAL, Univ. Paris-Sud and CNRS/IN2P3, Orsay, France\\
% ...
$^{118}$ Department of Physics, Oxford University, Oxford, United Kingdom\\
% ...
$^{153}$ Raymond and Beverly Sackler School of Physics and Astronomy, Tel Aviv University, Tel Aviv, Israel\\
% ...
\end{flushleft}


\end{document}